\documentclass[preprint,12pt,authoryear]{elsarticle}

\usepackage{amssymb}

\makeatother
\graphicspath{ {./figure/} }

\usepackage{alltt}
\usepackage{amsfonts}
\usepackage{amssymb}
\usepackage[tbtags]{amsmath}
 
\newtheorem{theorem}{Theorem}

\newtheorem{corollary}[theorem]{Corollary}

\newtheorem{result}[theorem]{Result}

\newenvironment{proof}[1][Proof]{\textbf{#1.} }{\ \rule{0.5em}{0.5em}}

\newcommand{\kb}[1]{\mathbf{#1}}

\newcommand{\kbm}[1]{\boldsymbol{#1}}

\def \R {\mathbb{R}}

\def \th {\theta}
\def \g {\gamma}

\def \l {\lambda}

\def \SO \mathcal{SO}
\def \Or \mathcal{O}

\def \ex \text{exp}

\newcommand{\intl}{\int\limits}
\newcommand{\prodl}{\prod \limits}
\usepackage{amsmath,amssymb,color,bm}

\definecolor{c20}{rgb}{0.,0.7,0.}
\definecolor{c30}{rgb}{0.,0.,.7}
\definecolor{c40}{rgb}{1,0.1,0.7}
\newcommand{\AK}[1]{\textcolor{c30}{\bf{#1}}}

\definecolor{c50}{rgb}{0,0,0}

\newcommand{\AKL}[1]{\textcolor{c40}{#1}}

\newcommand{\mybg}{\bm{\gamma}}

\newcommand{\mybth}{\bm{\theta}}

\newcommand{\Cth}{\mathcal{C}(\kbm{\th})}

\newcommand{\x}{\bm{x}}

\begin{document}

\begin{frontmatter}

\title{A branch cut approach to the probability density and distribution functions of a linear combination of central and non-central Chi-square random variables}
\author{Alfred Kume}
	\ead{a.kume@kent.ac.uk}
 \affiliation{organization={SMSAS} , 
 addressline={University of Kent},	
city={Canterbury},
country={U.K.}}

 	\author{Tomonari Sei}
 \ead{sei@mist.i.u-tokyo.ac.jp}
 	\affiliation{organization={The University of Tokyo} , 
 		city={ Tokyo},
 		country={Japan}}

\author{Andrew T. A. Wood}
\ead{ andrew.wood@anu.edu.au}
\affiliation{organization={Research School of Finance, Actuarial Studies \& Statistics} , 
	addressline={Australian National University},	
	country={Australia}}





\begin{abstract}
The paper considers the distribution of a general linear combination of central and non-central chi-square random variables by exploring the branch cut regions that appear in the standard Laplace inversion process. Due to the original interest from the directional statistics, the focus of this paper is on the density function of such distributions and not on their cumulative distribution function. In fact, our results confirm that the latter is a special case of the former. Our approach provides new insight by generating alternative characterizations of the probability density function in terms of a finite number of feasible univariate integrals. In particular, the central cases seem to allow an interesting representation in terms of the branch cuts,  while general degrees of freedom and non-centrality can be easily adopted using recursive differentiation. 
Numerical results confirm that the proposed approach works well while more transparency and therefore easier control in the accuracy is ensured.
 
\end{abstract}

\begin{keyword}
	Linear combination of Chi-squares\sep Bingham distributions\sep Fisher-Bingham distributions\sep directional statistics\sep holonomic functions.
	
	
\end{keyword}

\end{frontmatter}

\section{Introduction}

Let us denote by $X$ a  quadratic form of normally distributed components: 
\begin{equation}
	X=\sum_{i=1}^{p} Y_i^\top Y_{i}=\sum_{i=1}^{p}
	\lambda_i\chi^2_{n_i}(\delta_i)\quad \lambda_i=\frac{1}{2\theta_i} \quad  \delta_i=\frac{\gamma_i^2}{2 \th_i}
	\label{eqn:gen:rep}
\end{equation}
where $Y_i \sim N_{n_i}(\frac{\gamma_i}{2 \theta_i \sqrt{n_i}}\textbf{1},\frac{1}{2\theta_i}\textbf{I}_{n_i})$ are $p$ independent multivariate normal rv's of dimension $n_i$, each with covariance matrix the multiple of identity $\frac{1}{2\theta_i}\textbf{I}_{n_i}$ and mean some vector of norm $\frac{\gamma_i}{2 \theta_i}$. 

Evaluating the probability density function (pdf)  and the cumulative distribution function (cdf) of such quadratic forms has important applications in statistical theory. Many statistical tests such as the goodness of fit for non-parametric regression (e.g. \cite{Kuonen2005}), pseudo-likelihood ratio tests (e.g. \cite{Liang1996}), directional statistics models  using the Fisher-Bingham distributions (see \cite{MardiaJupp2000,KW05}) or the shape distributions based on the complex Bingham family (see  \cite{Kent1994}), are heavily relying on data modelled by these types of random variables. Since in their most general cases there is no close form expression for the corresponding distributions the numerical evaluation is implemented  in many situations.  
Representation \eqref{eqn:gen:rep} is implicitly assuming that $\th_i$ as well as $\lambda_i$ are positive.
The more general cases when some $\th_{i}$'s are negative can be seen as an extension of \eqref{eqn:gen:rep} since by collecting the terms according to their signs,  it can be considered as a difference of two such positive combinations.  We will consider these cases separately. 


Standard inverting arguments of the corresponding pdf's  via the moment generating function or Laplace transform confirm that
the density function of $X$ is  
\begin{eqnarray}
f_{\mybth, \mybg^2, \kbm{n}}(s)&=& 
\prod_{i=1}^{p} \th_i^{\frac{n_i}{2}} \exp (-\sum_{i=1}^{p}\frac{\gamma^2_i}{4\th_i}) \frac{1}{ 2\pi \textbf{i}} \intl_{\textbf{i}\R+t_0}  \frac{\text{exp}(\sum_{i=1}^p \frac{\gamma_i^2}{4(\th_i+t)})}{\prod_{i=1}^p(\th_i+t)^{\frac{n_i}{2}} }   e ^{st} dt, \quad s>0
\label{eqn:dens}
\end{eqnarray}
where $t_0$ can be any real number such that the vertical line ${\textbf{i}\R+t_0}$ in the complex plane is on the right of the poles $-\th_i$ i.e. $-\min(\mybth)<t_0$. Note that throughout the paper we refer to $\mybth, \mybg^2, \kbm{n}$ as the vectorised entries of $\th_i$, $\gamma_i$ and $n_i$ respectively where $\th_i$ are forced to be distinct; and we apply the usual convention for complex integration so that for the closed contours we integrate along the anticlockwise direction and for the unbounded lines such as ${\textbf{i}\R+t_0}$ we integrate from $t_0-{\textbf{i}\infty}$ to $t_0+{\textbf{i}\infty}$. 
The current methods addressing this inversion problem could be categorised into  three groups: 

\textit{Standard numerical inversion}: 
Attempts to numerically evaluate these functions go back as far as in \cite{IMH:61} where a general numerical integrating procedure along a vertical line similar to that in \eqref{eqn:dens} is shown. 
The alternative approach introduced in \cite{Davies1973} and \cite{Davies1980} is addressing the evaluation of the cumulative distribution function calculation with numerical performance comparable to that of \cite{IMH:61}. 
A moment matching procedure to some gamma distribution is reported in \cite{Liu2009}. However as indicated by \cite{Duchesne2010} this approach is not always guaranteeing good performance. 
All of the above mentioned  methods are generally focused on the cdf's and not on the pdf $f_{\mybth, \mybg^2, \kbm{n}}$.



\textit{Saddle point approximation} (SPA): This method relies on the extensions of the Laplace approximation for integrals (see e.g. \cite{Daniels1954,BarndorffNielsen1990a}). The manuscript of  \cite{Butler2007} is an  excellent review of this popular method. Its implementation in our  context is straightforward with  little computational cost.  This approach is  shown in \cite{Kuonen1999} to perform well with a surprising accuracy in non-parametric regression examples reported. 
  The same conclusion is reported by \cite{KW05} in the context of the density function evaluation. The authors implemented this for directional statistics inference involving the evaluation of  $f_{\mybth, \mybg^2, \kbm{n}}$ at point $s=1$. 

\textit{Holonomic gradient methods}: The evaluation of $f_{\mybth, \mybg^2, \kbm{n}}$ is in fact characterized as a solution of a particular ODE equation (c.f. \cite{hgd}) where in principle  only some accurate starting point and a good ODE solver is required. The derivation of such ODE derives from the fact that  $f_{\mybth, \mybg^2, \kbm{n}}$ is shown to be an holonomic function for which the Pfaffian matrix equations lead to the relevant ODE. The general adjustments for allowing multiplicities in $\mybth$ so that the corresponding ODE do not become degenerate, are  studied in \cite{Sei2013,Kume2018} for central and non central cases respectively. In fact \cite{Koyama2016} focus on the problem of using HGM for the cdf. 

This paper belongs to the first group of these approaches. We focus on exploring the specific structure of the Laplace transform so that a more flexible approach both numerically and algebraically is offered.
More specifically, the main contribution aspects  of this paper are: 
\begin{itemize}
	\item Theorem 1 expresses the integrating contour in a much simpler form as seen before in the literature, leading to numerically simpler inversion procedures.  This is achieved by exploring the simple branch cut structure due to the square-root functions.  Additionally, we establish a close connection between the cdf of $X$ as  in \eqref{eqn:gen:rep} and the pdf of some other positive linear combination of non-central Chi-squares $X'$ such that
	\begin{equation}
		P(X\le x)=f_{\tilde{X}}(x)\frac{e ^{x\th_0}}{\th_0} \prod_{i=1}^{p}\left( \frac{\theta_i}{\theta_i+\th_0}\right)^\frac{n_i}{2} 
		\text{exp}(-\sum_{i=1}^{p}\frac{\gamma^2_i \th_0}{4\th_i(\th_i+\th_0)})\quad \forall \th_0>0
		\label{eqn:pdf:cdf}
	\end{equation}
	where $f_{\tilde{X}}$ is the pdf of $ \tilde{X}=	\frac{1}{2\theta_0} \chi^2_2(0)+\sum_{i=1}^{p}
	\frac{1}{2(\theta_i+\theta_0)} \chi^2_{n_i}(\frac{\gamma_i^2}{2 (\th_i+\th_0)})$ for any arbitrary $\th_0$.
	An immediate consequence of this  fact is that any general method for evaluating the pdf  of any positive linear combination of non-central Chi-square random variables suffices for the distribution functions too. For example the saddlepoint approximation of pdf's as in \cite{KW05} can also be used for the tail probabilities as an alternative to that of \cite{Kuonen1999,BarndorffNielsen1990a}. Additionally, the ODE equations of \cite{Koyama2016} for cdf. could also be seen as a special case of those developed in \cite{Sei2013,Kume2018} for the pdf. 
	To our knowledge property \eqref{eqn:pdf:cdf} has not been reported in the literature before. 
	\item Theorem 2 makes the corresponding simplifications for the practically important central cases, $\kbm{\g}=0$, where the univariate contours are now reduced to a finite linear combination of some bounded elementary univariate integrals which are numerically manageable and therefore the normalising constants of the Bingham distributions can be easily derived. 
	\item Theorem 3 extends the above mentioned  results to the difference of two positive linear combinations  including the corresponding result as in \eqref{eqn:pdf:cdf}. 
	In particular, for the case of chi-square distributions of degrees of freedom 2 which corresponds to the rescaled exponential distributions, a closed form expression is immediately  obtained using residues of simple poles. 
	\item The relative simplicity and efficiency of our inversion is confirmed through numerical examples. In particular, we show that we can not  just simply offer individual entries to the tail probabilities like shown before but the whole function is easily evaluated due to feasibility of the numerical integration. 
	We have successfully implemented here the standard numerical integration routines within the R package where, if required,  the multiple precision integration is available calling the GMP library  of the C programming environment. 
\end{itemize}

The Laplace inversion \eqref{eqn:dens} provides additional identities for these distributions.  More specifically, using the flexibility on the choice of such $t_0$ and applying a change of variable $u=s(t+c)$, for any real value of $c$
implies the rescale and shift property 
\begin{equation}
f_{\mybth, \mybg^2, \kbm{n}}(s)=\frac{e^{-sc}}{s}f_{s\mybth-sc,s\mybg^2}(1)\prod_{i=1}^{p}\frac{\th_i^{\frac{n_i}{2} }}{(\theta_i-c)^{\frac{n_i}{2}}}  \text{exp}(\sum_{i=1}^{p}\frac{\gamma^2_i}{4(\th_i-c)}-\frac{\gamma^2_i}{4\th_i}).
\label{eqn:property:rescale}
\end{equation}

In fact, $f_{s\mybth-sc,s\mybg^2}(1)$
is  also related to that of evaluating the normalising constant of the Fisher-Bingham distributions on $\sum_{i=1}^{p} n_i-1$ dimensional spheres 
(c.f. \cite{KW05}): 
\begin{equation}
	\mathcal{C}(\kbm{\th },\kbm{\g}, \kbm{n})
	=\intl_{\mathcal{S}^{\sum_{i=1}^{p} n_i-1}}  \text{exp}( \sum_{i=1}^p -\th_i x_i^2+\g_i x_i) d_{\mathcal{S}^{p-1}}(\x) 
	=2 \pi^{\frac{\sum_{i=1}^{p} n_i}{2}}    \frac{ \text{exp}( \sum_{i=1}^p \frac{\gamma^2_i}{4 \th_i })}{\prod_{i=1}^p\th_i^{\frac{n_i}{2}}}f_{\mybth,\mybg^2, \kbm{n}}(1)
	\label{eqn:dens:FBnc}
\end{equation}where $d_{\mathcal{S}^{\sum_{i=1}^{p} n_i-1}}(\x)$ is the uniform measure   
and if all $\gamma_i=0$, $\mathcal{C}(\kbm{\th },\kbm{0},\kbm{n})=\mathcal{C}(\kbm{\th },\kbm{n})$ reduces to that of the Bingham distribution.
The relationship
\begin{equation}
	f_{\mybth, \mybg^2,\kbm{n}}(s)=\frac{ \prod_{i=1}^p\th_i^{\frac{n_i}{2}} \text{exp}( \sum_{i=1}^p -\frac{\gamma^2_i}{4 \th_i })}{2 \pi^{\frac{\sum_{i=1}^{p} n_i}{2}}  }  s^{\frac{\sum_{i=1}^{p} n_i}{2} -1} \mathcal{C}(s\kbm{\th },\sqrt{s}\kbm{\g},\kbm{n})\label{eqn:dens:sph_nc}
\end{equation}

implies that the methods developed for evaluating the normalizing constants for  directional distributions  apply immediately to those of the density function of some linear combination of chi-squares. In fact the shift property \eqref{eqn:property:rescale} implies that $\th_{i}$ in \eqref{eqn:dens:FBnc} can be allowed to be shifted at negative values, as it is the case for alternative parametrisations of $ \mathcal{C}(\kbm{\th },\kbm{\g},\kbm{n}) $ where $\theta_{i}$ can be either positive or negative. 
An additional property which derives from \eqref{eqn:dens} is the connection between multiplicities (or degrees of freedom) $n_i$ and differentiation:
\begin{equation}
	f_{\mybth, \mybg^2,\kbm{n}}(s)\prod_{i=1}^{p} \th_i^{-\frac{n_i}{2}} e^{-\sum_{i=1}^{p}\frac{\gamma^2_i}{4\th_i}}=\mathcal{D}\mathcal{D'} \left(f_{\mybth^0, \mybg^2,\kbm{n}^0}(s)\prod_{i=1}^{p} \th_i^{-\frac{n^0_i}{2}} \text{exp}(-\sum_{i=1}^{p}\frac{\gamma^2_i}{4\th_i})\right)
	\label{eqn:diff:property}
\end{equation}
where 
$$ \mathcal{D'}=\prod_{i=1}^{p_1}\frac{(-1)^{r_i}}{n^0_i(n^0_i+1)\cdots(n^0_i+r_i-1)}\frac{\partial^{r_i}}{\partial^{r_i} (\th'_i)}
\quad\text{}\quad \mathcal{D}=\prod_{i=1}^{p-p_1}\frac{\partial^{r_i}}{\partial^{r_i}(\g_i^2/4)}
$$
and $f_{\mybth^0, \mybg^2,\kbm{n}^0}(s)$ representing the pdf  with each $\th_i$ having multiplicity $n^0_{i}\in \{1,2\}$,  $n_i=2r_i+n^0_i$ and  the first $p_1$ terms are collected to represent the central cases i.e. $\g_i=0$.

These differentiation rules generalize similar results of \cite{KumeWood2007} for the normalizing constants of Bingham distributions and those of \cite{IMH:61} for the pdf.

While there have been ongoing  attempts in the literature (see the recent work in \citet{Chen2021} and the references therein), to address this problem numerically,   they are mostly focused only in the case of positive $\theta_{i}$. In fact,  the properties reported here imply alternative computation methods which are easily implemented terms of some elementary integrating terms and  also hold even if  in  \eqref{eqn:dens} some $\th_i\le 0$ . 
Therefore if  extreme accuracy is required,  standard multiple precision integration can be easily  adopted to the resulting elementary functions.

The paper is organized as follows. In Section~2 we explore the branch cuts of the integrand function of \eqref{eqn:dens} and then adjust the contour accordingly. This gives rise to the connection of the pdf  expressions with those for the cdf stated in  \eqref{eqn:pdf:cdf}. In the following section we report the special adaptations for the relevant quantities for practical implementation. We then extend in  Section~4 the results to  that of a difference between two quadratic forms  of type \eqref{eqn:gen:rep}. We report some numerical illustrations in  Section~5  and conclude with some general remarks. 



\section{Exploring the branch cut structure}

We focus initially on the complex valued function
\begin{equation}
h(t)=\frac{1}{\sqrt{(\th_1+t)(\th_2+t)\cdots(\th_p+t)}}=\prod_{i=1}^p r_i^{-1} e^{\text{-\textbf{i}} \alpha} \quad \alpha=\sum_{i=1}^p \alpha_i
\label{eqn:alp}
\end{equation}
which corresponds to the central case, $\kbm{\gamma}=\kbm{0}$.
For each term above we use the parametrization 
$$\sqrt{\th_i+t}=r_i e^{\text{\textbf{i}} \alpha_i } \quad \text{where} \quad \alpha_i \in (+\pi, -\pi).$$ It is important to note that these terms are individually not analytic  for each positive $r_i$ and $\alpha_{i}=\pm \pi$
and  the corresponding branch cut is  the line of real numbers on the left of $-\th_i$. The product of square roots however leads to pairwise branch cuts as 
$$
C_r=[-\th_{2r},-\th_{2r-1}]
$$
which are segments determined by pairs of $-\th_i$. If $p$ is odd then the last value $-\th_p$ will be paired with $-\infty$ as 
$$
C^{\infty}_{[\frac{p+1}{2}]}=[-\th_{p},-\infty].
$$
One can easily see that in general the product
$$
\frac{1}{\sqrt{(\th_1+t)(\th_2+t)\cdots(\th_p+t)}}
$$
has $k=[\frac{p+1}{2}]$ branch  cuts
$$
C_1,C_2\cdots C_k 
$$
such that if $p=2k-1$, $C_{k+1}$ is unbounded.
 
Please note that the choice of these branch cuts is somewhat arbitrary as each square root term can allow branches on either side of the respective pole $-\th_{i}$ and therefore each pair of products $\sqrt{(\th_i+t)(\th_j+t)}$ can generate either an individual branch cut as the segment $[\th_{i},\th_{j}]$ or its complement in the real line. More specifically, using a different parametrization, one could alternatively generate the unbounded branch cut extending from $-\th_1$ to $+\infty$ instead.

\begin{figure}[ht!]
 	\begin{minipage}[b]{0.45\linewidth}
 		\centering
 		\centering
 		\includegraphics[scale=.1]{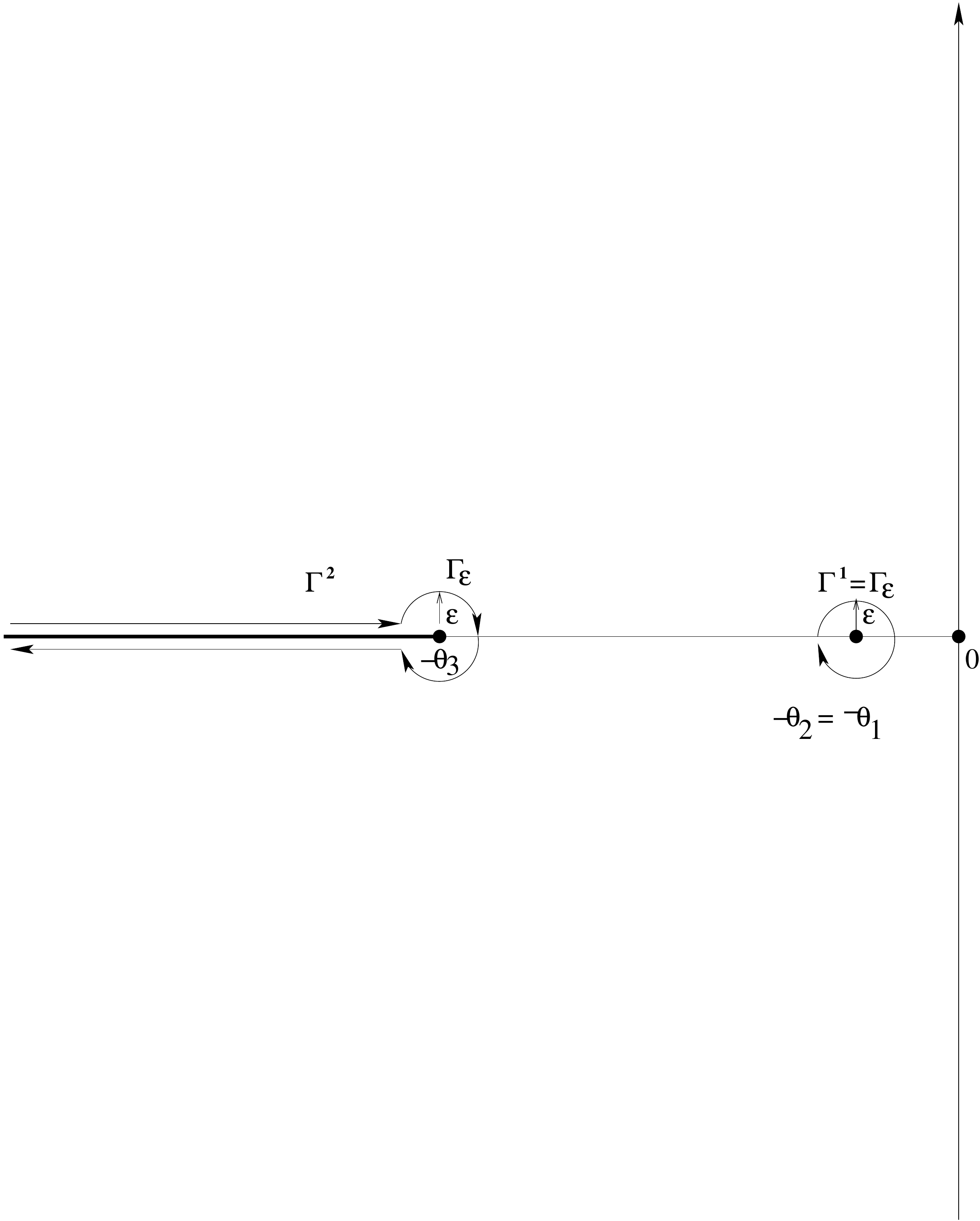} 
 		\caption{Contour limit as $\th_{2}\to\th_{1}$ for $p=3$.}
 		\label{fig:even_1}
 	\end{minipage}
 	\hspace{0.01cm}
 	\begin{minipage}[b]{0.45\linewidth}
 		\centering
 		\includegraphics[scale=.1]{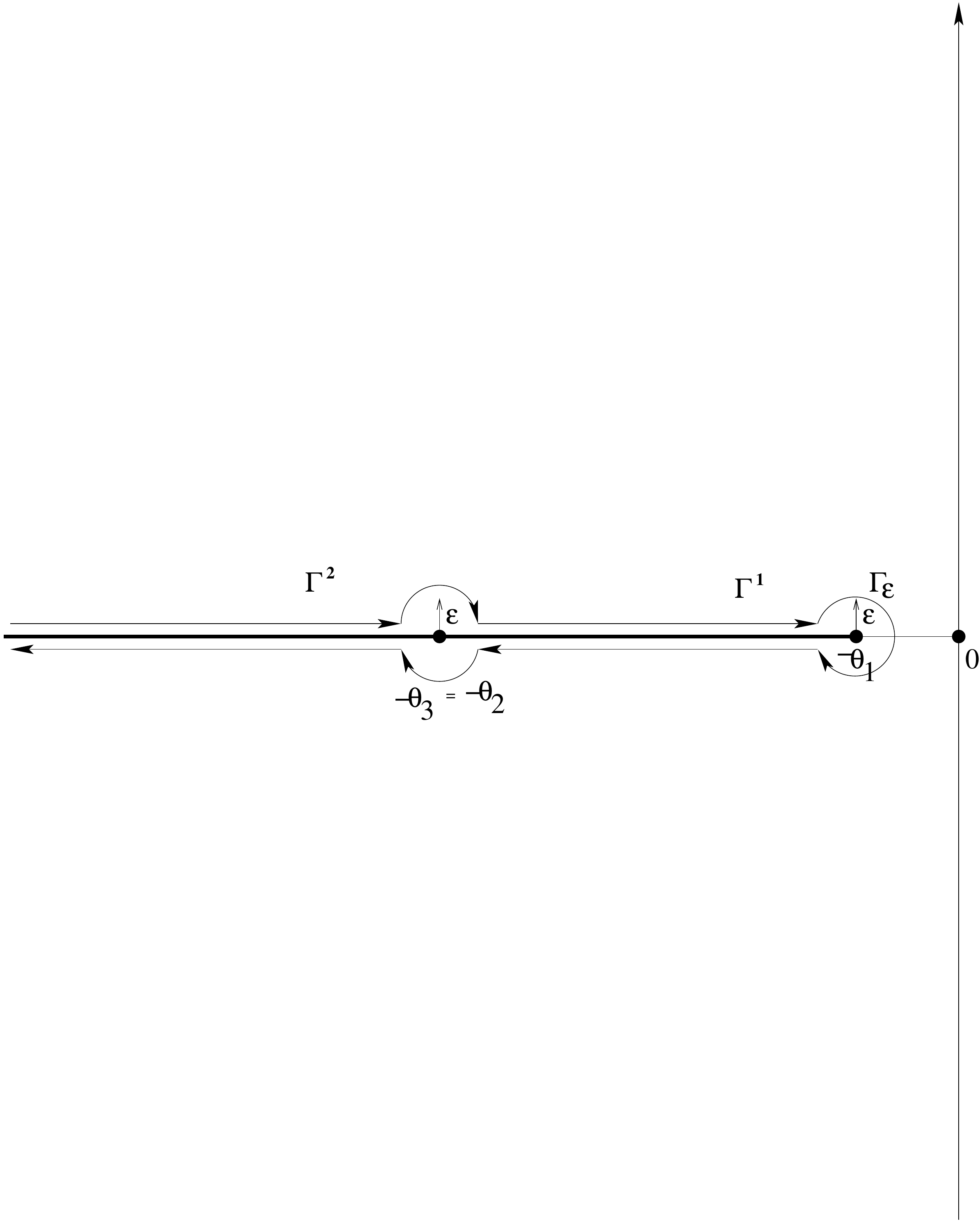} 
 		\caption{Contour limit as $\th_{2}\to\th_{3}$ for $p=3$.}
 		\label{fig:even_2}
 	\end{minipage}
 \end{figure}

 We are using here the ordered pairs so that the interpretation is meaningful such that it allows for the odd number $p$ to generate the leftmost branch extending to $-\infty$. 
 Since outside these branch cuts the non-central components are analytical functions except the respective  poles $-\th_{i}$ then the branch cut structure is not affected also for
 $$
g(t)=\prod_{i=1}^{p}\frac{\text{exp}(\sum_{i=1}^p \frac{\gamma_i^2}{4(\th_i+t)})}{(\th_i+t)^{k_i}} \frac{1}{\prod_{i=1}^p\sqrt{\th_i+t} }=\prod_{i=1}^{p}\frac{\text{exp}(\sum_{i=1}^p \frac{\gamma_i^2}{4(\th_i+t)})}{(\th_i+t)^{k_i}} h(t)
 $$
 which corresponds also to the cases of \textbf{odd} degrees of freedom, i.e. $n_i=2k_i+1$. 
 If we have \textbf{even multiplicities} for some particular  $\th_{i}$, say $n_{i}=2k_{i}$ then  there is no presence in the $h(t)$ part of the integrand and the contribution of the corresponding term $\frac{1}{(\th_{i}+t)^{k_{i}}}$ is  analytical. Therefore  the branch cut structure is not affected except for the appropriate degeneracy at the respective pole at $-\th_i$.  
 In fact, an even multiplicity can appear in two types:
 \begin{enumerate}
 	\item[\textbf{Type 1}] appears if the branch cut $C_r$ reduces to a single point i.e. $\th_{2r}=\theta_{2r-1}$ (see the pole $-\th_1=-\th_2$ in Figure \eqref{fig:even_1}, for $p=3$ and $r=1$). This occurs in general if the number of distinct $\th_{i}$ less than $\theta_{2r-1}$ is even.  
 	\item[\textbf{Type 2}] appears if two adjacent branch cuts merge so that the joining extreme appears with multiplicity 2 (see the pole $-\th_3=-\th_2$ in  Figure \eqref{fig:even_2}). 
 \end{enumerate}
 
 \begin{theorem}
 	Assume that $p$ distinct and increasingly ordered $\th_{i}$ have multiplicities $n_{i}$  generating $r$ disjoint segments of branch cuts $C_{k}$ and $l$ particular poles of even multiplicities  of type 1.  
 	Then the corresponding pdf and cdf expressions of the random variable $X=\sum_{i=1}^{p}
 	\frac{1}{2\theta_i}\chi^2_{n_i}(\frac{\gamma_i^2}{2 \th_i})
 	$ are
 	$$	f_{\mybth, \mybg^2,\kbm{n}}(s)=\frac{f_{\mybth s, \mybg^2 s,\kbm{n}}(1)}{s}=\kappa \left(\frac{-1}{2 \pi\textbf{i}} \sum_{i=1}^{r} \intl_{\Gamma^{i}}   g(t) e^{st}  dt +\frac{-1}{2 \pi\textbf{i}} \sum_{j=1}^{l} \intl_{\Gamma^{j}_{\varepsilon}}   g(t) e^{st}  dt\right) $$	
 	where 	$$
 	g(t)=\frac{\exp(\sum_{i=1}^p \frac{\gamma_i^2}{4(\th_i+t)})}{\prod_{i=1}^p(\th_i+t)^{\frac{n_i}{2}} } \quad \kappa=\prod_{i=1}^{p} \theta_i^{\frac{n_i}{2}} \exp(-\sum_{i=1}^{p}\frac{\gamma^2_i}{4\th_i})  $$
 	and $\Gamma^{i}$ are the non overlapping contours enveloping the branch cut regions $C_{i}$ determined by $\th_{i}$ while $\Gamma^{j}_{\varepsilon}$ are the non overlapping contours around the points with even multiplicities $\th_{j}$ of type 1. 
 		Additionally,
 		\begin{eqnarray*}
 			P(X\le x)&=& \frac{e ^{x\th_0}}{\th_0} f_{\tilde{\mybth},\tilde{\mybg}^2}(x) \prod_{i=1}^{p}\frac{\sqrt{\theta_i}}{\sqrt{\theta_i+\th_0}}
 			\exp(-\sum_{i=1}^{p}\frac{\gamma^2_i \th_0}{4\th_i(\th_i+\th_0)})\\
 			&=&
 			\frac{e ^{x\th_0}}{x \th_0} f_{x\tilde{\mybth},x\tilde{\mybg}^2}(1) \prod_{i=1}^{p}\frac{\sqrt{\theta_i}}{\sqrt{\theta_i+\th_0}}
 			\exp(-\sum_{i=1}^{p}\frac{\gamma^2_i \th_0}{4\th_i(\th_i+\th_0)})
 		\end{eqnarray*}
 		
 			where $f_{\tilde{\mybth},\tilde{\mybg}^2}$ is the density function of the  linear combination $ \tilde{X}=\frac{1}{2\theta_0} \chi^2_2(0)+\sum_{i=1}^{p}
 	\frac{1}{2(\theta_i+\theta_0)} \chi^2_{n_i}(\frac{\gamma_i^2}{2 (\th_i+\th_0)}), \forall \th_0>0$.
 \end{theorem}
 
 \textbf{The proof is presented in Appendix.}

 
%
 
\begin{figure}[ht]
	\begin{minipage}[b]{0.45\linewidth}
		\centering
		\centering
		\includegraphics[scale=.12]{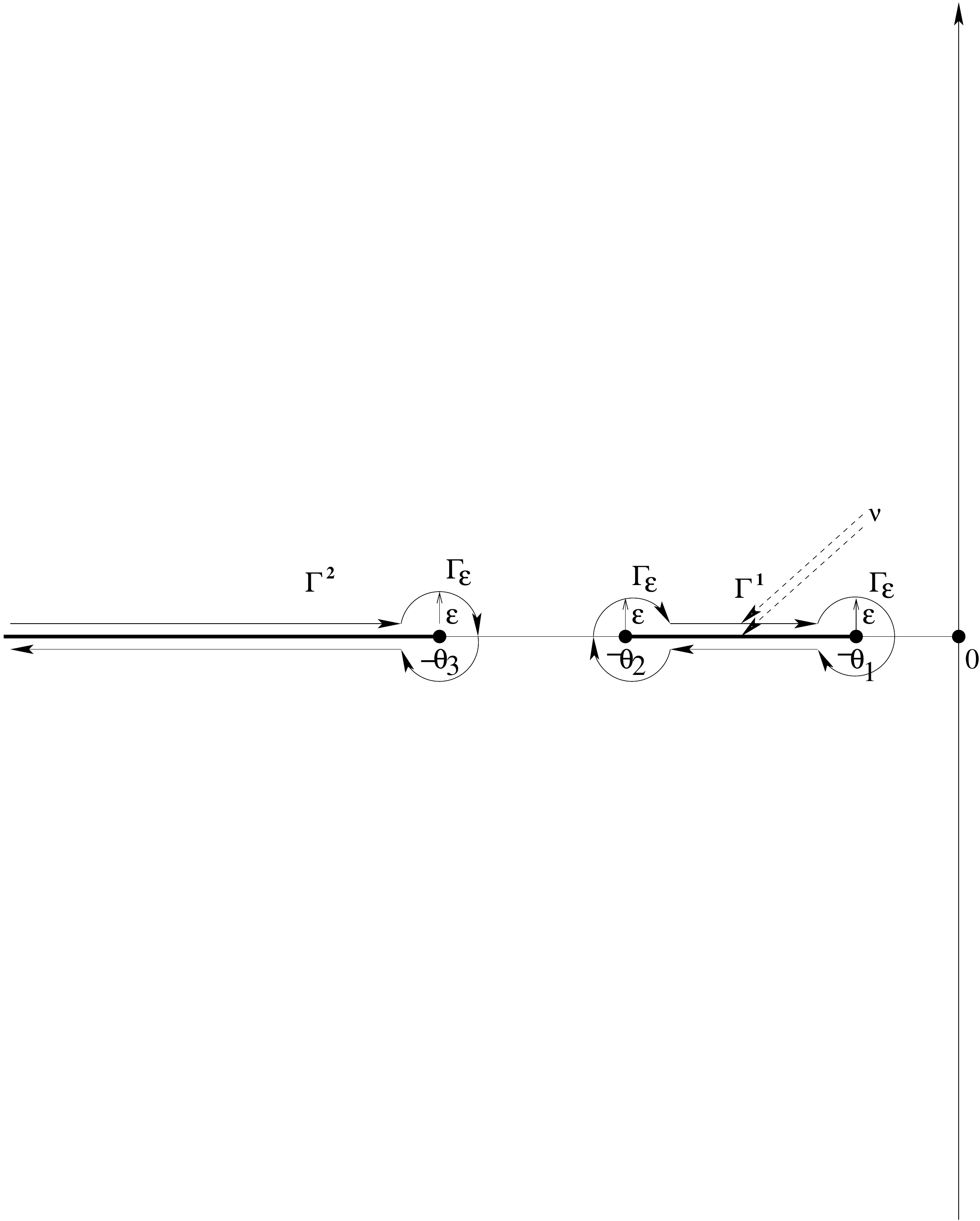} 
		\caption{Contours for non central cases}
		\label{fig:simple:gen}
	\end{minipage}
	\hspace{0.01cm}
	\begin{minipage}[b]{0.45\linewidth}
		\centering
		\includegraphics[scale=.12]{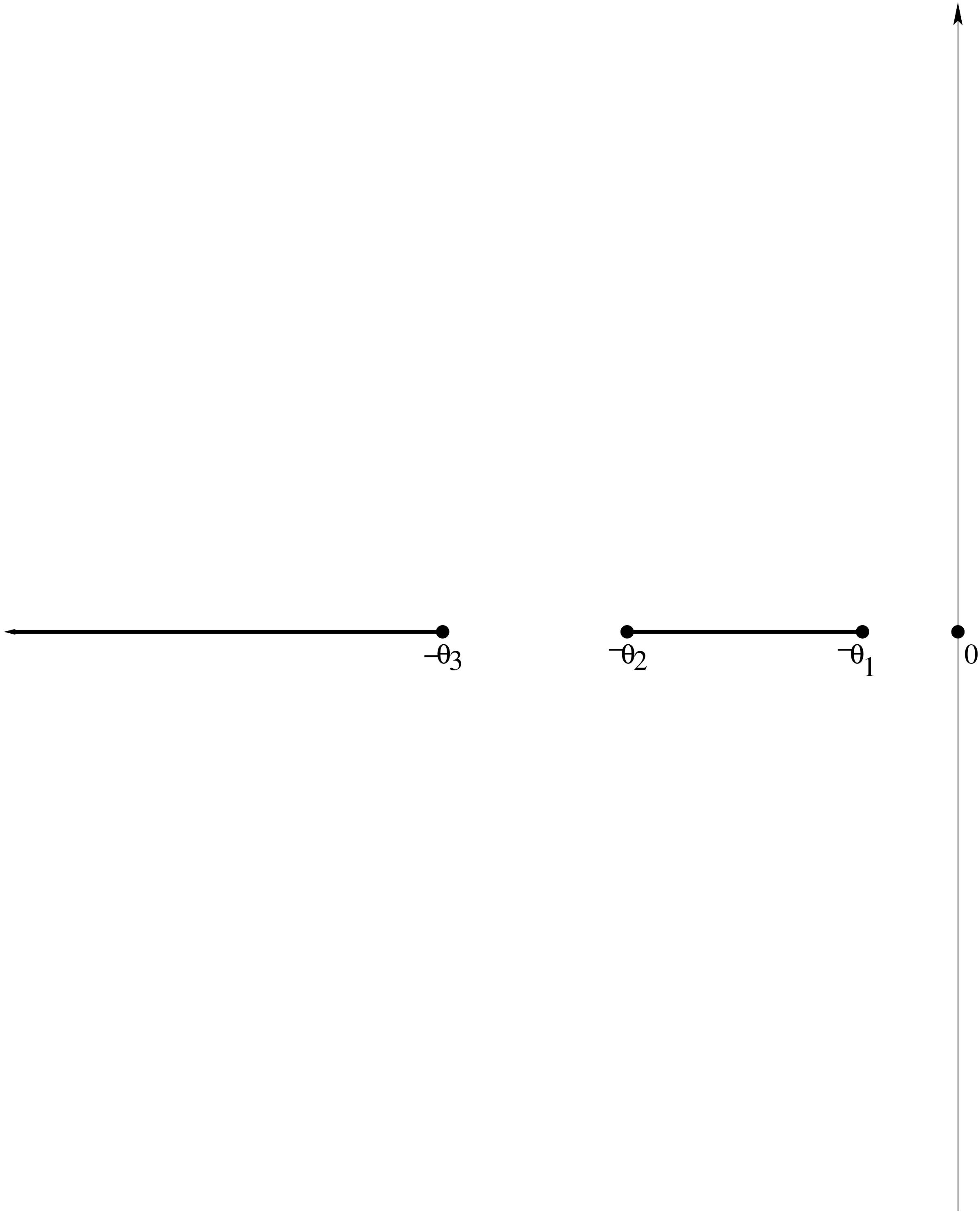} 
		\caption{Contour limit as $\nu,\varepsilon\to 0$}
		\label{fig:simple:lim}
	\end{minipage}
\end{figure}

 Theorem~1 suggests that it is sufficient to integrate along the non-degenerate contours like ``dog-bone" type as $\Gamma^{1}$ or ``keyhole" type as $\Gamma^{2}$ around branch cuts  $C_{1}$ and $C_{2}$, see Figure \eqref{fig:simple:gen}.  Integration along such curves, if appropriately parametrized, gives rise to univariate integrals. For example, the relevant term here $\frac{-1}{2 \pi\textbf{i}} \intl_{\Gamma^{i}}   g(t) e^{st}  dt$ is only the imaginary part of the integrand so that one could use  $\frac{-1}{2 \pi} \intl_{\Gamma^{i}}   \textbf{Im}(g(t) e^{st})  dt$ instead. In our practical implementation reported in Section~5, we use the Romberg integration method on spherical contours which seems to work well within the R package which can be easily utilised to use the multiple precision routines. If more control or  efficiency is required other lower level programming tools can in principle be used. 
 
 In order to obtain the contours $\Gamma^{i}$ and $\Gamma^{j}_{\varepsilon}$ of Theorem above, for a given set of $p$ distinct $\th_{i}$ with multiplicities $n_{i}$, one could perform the following two steps:
 \begin{enumerate}
 	\item Consider initially only those $\th_{i}$ which have odd multiplicities and pair them to determine the corresponding branch cuts $C_{i}$  as in   $\Gamma^{i}$'s of Theorem. 
 	\item Consider next only those $\th_{i}$ with even multiplicities(type 1) which are outside $\Gamma^{i}$ regions and calculate their residues either numerically by integrating along contours $\Gamma^{j}_{\varepsilon}$ or analytically if $\gamma_i=0$.
 \end{enumerate}
 
In fact the integrating contours reduce to line segments for the practically important case  of all terms being non-central chi squares with single degrees of freedom,  namely,  the case of $p$ distinct $\theta_{i}$, each having multiplicity $n_i=1$ and $\gamma_i=0$. 
 
\begin{theorem}
For distinct $\th_{i}>0$ and $\g_i=0$, the corresponding pdf and cdf expressions for the random  variable $X=\sum_{i=1}^{p}
\frac{1}{2\theta_i}\chi^2_{1}(0)
$  are:
	\begin{eqnarray}
	f_{\mybth, \kbm{0},\kbm{1}}(s)&=& 
	\frac{\prod_{i=1}^{p}\sqrt{\theta_{i}}}{\pi} \left( \int_{\theta_1}^{\theta_2}  \frac{e^{-st}}{ \sqrt{-\prod_{i=1}^p (\th_i-t)}} dt-\int_{\theta_3}^{\theta_4}  \frac{e^{-st}}{ \sqrt{-\prod_{i=1}^p (\th_i-t)}} dt+\int_{\theta_5}^{\theta_6}  \frac{e^{-st}}{ \sqrt{-\prod_{i=1}^p (\th_i-t)}} dt  \cdots\right)\nonumber\\
	&=& \frac{\prod_{i=1}^{p}\sqrt{\theta_{i}}}{\pi} \sum_{r=1}^{[\frac{p+1}{2}]}(-1)^{r+1} \int_{\theta_{2r-1}}^{\theta_{2r}}  \frac{e^{-st}}{\sqrt{-\prod_{i=1}^p (\th_i-t)}} dt
	\end{eqnarray}
where $$\theta_{2[\frac{p+1}{2}]}=\left \{ \begin{array}{cc}
\theta_{p}& \text{p even}\\
\infty& \text{p odd} 
\end{array} \right. $$
and 
$$
P(X\le x)=
f_{\tilde{\mybth}, \kbm{0},\kbm{1}}(x)\frac{e^{x\th_0} \prodl_{i=1}^p\sqrt{\th_i}}{\th_0 \prod_{i=1}^{p}\sqrt{\theta_{i}+\th_0}}=f_{\tilde{\mybth}x, \kbm{0},\kbm{1}}(1)\frac{e^{x\th_0} \prodl_{i=1}^p\sqrt{\th_i}}{ x\th_0 \prod_{i=1}^{p}\sqrt{\theta_{i}+\th_0}}\quad \forall \th_0>0
$$
where $f_{\tilde{\mybth}, \kbm{0},\kbm{1}}$ is the pdf of $\tilde{X}=	\frac{1}{2\theta_0} \chi^2_2(0)+\sum_{i=1}^{p}
\frac{1}{2(\theta_i+\theta_0)} \chi^2_{n_i}(0)$.
\label{Theo:Bingh}	
\end{theorem}
\textbf{The proof is presented in Appendix.}

In fact these theorems indicate that the regions where the relevant quantities need to be integrated are only those around the branch cuts appearing from $\th$'s with odd multiplicities (or chi-squared with odd degrees of freedom). In the central cases as in Theorem~2, these regions can be also reduced to simple lines. 
Contour integration is also valid along any lines containing all these branch cut regions and in fact any such choice will lead to a different integrating parametrization.

\subsection{Computation remarks for the central cases $\g_i=0$} 
\begin{figure}
	\centering
	\includegraphics[width=0.3\linewidth]{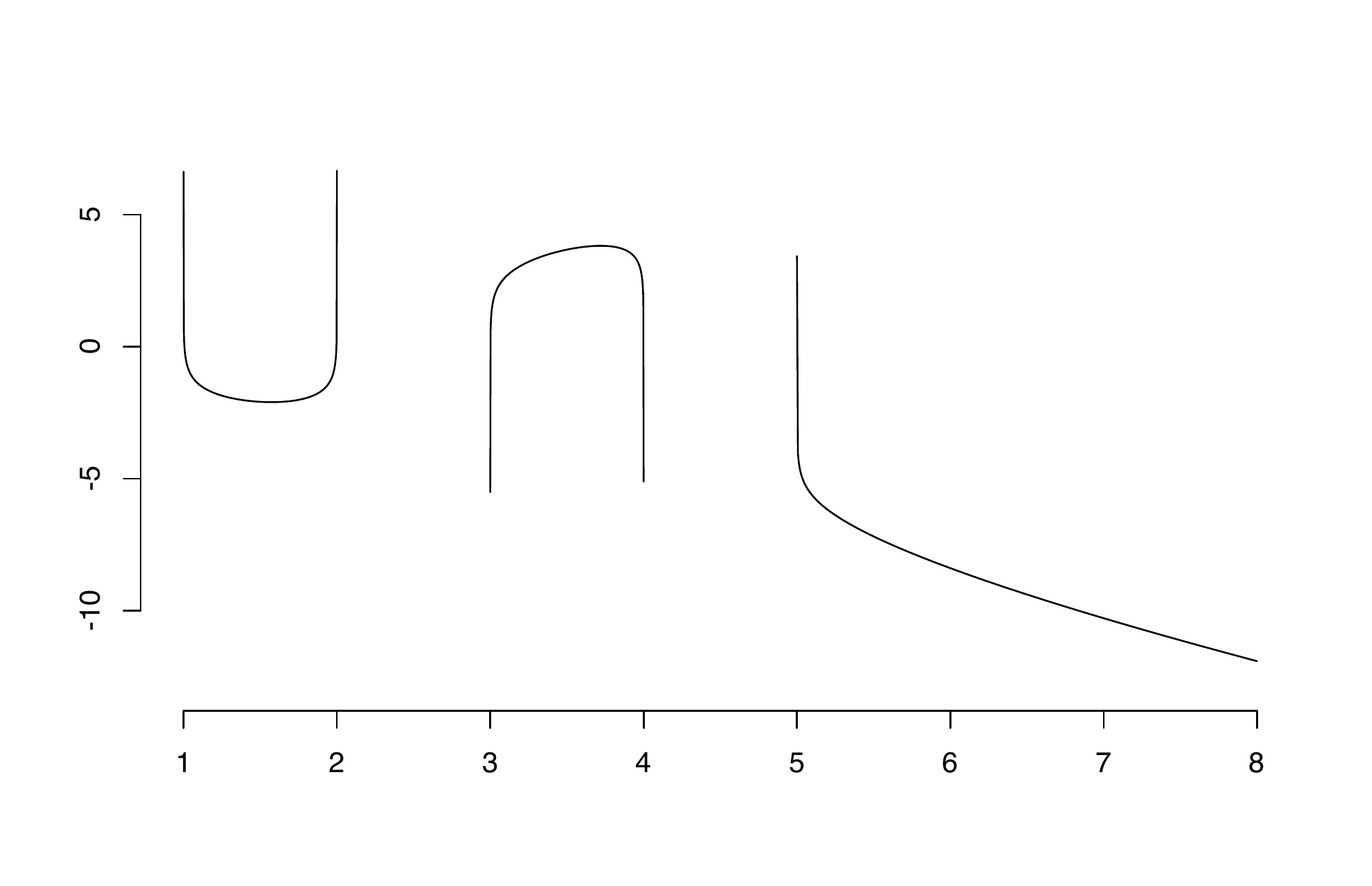}
	\caption{Function $g(t)e^{-st}$ in log-scale for the central case with $\kbm{\th}=c(1,2,3,4,5)$ and $s=1$.}
	\label{fig:g}
\end{figure}

The univariate integrating terms of Theorem~\ref{Theo:Bingh} are easily seen to be finite while can have alternative representation: 
\begin{eqnarray}
\intl_{\th_{2r-1}}^{\th_{2r}}  \frac{e^{-st}}{ \sqrt{-\prod_{i=1}^p(\th_i-t)} }  dt&=&e^{-s\th_{2r-1}}
\int_0^1 \frac{e^{-su\alpha_r} du}{\sqrt{u(1-u)\prodl_{2r-1\ne i\ne 2r}^p(\th_i-\th_{2r-1}-u\alpha_r)}} \label{eqn:uni:repar}
\\
&=&
2 e^{-s\th_{2r-1}}
\int_0^{\pi/2} \frac{\text{exp}(-s\alpha_r\sin^2u) du}{\sqrt{\prodl_{2r-1\ne i\ne 2r}^p(\th_i-\th_{2r-1}\cos^2 u-\th_{2r}\sin^2 u)}}
\nonumber
\end{eqnarray}
where $\alpha_r=\th_{2r}-\th_{2r-1}$ and $t=\th_{2r-1}\cos^2 u+\th_{2r}\sin^2u$. In the type one multiplicity case i.e.~$\alpha_{r}=0$, one can progress by taking the limit on \eqref{eqn:uni:repar} as $\alpha_r\to 0$:
\begin{equation}
\intl_{\th_{2r-1}}^{\th_{2r}}  \frac{e^{-st}}{ \sqrt{-\prod_{i=1}^p(\th_i-t)} }  dt=\xrightarrow[\alpha_r\to 0]{} \frac{\pi e^{-s\th_{2r-1}}}{\prodl_{2r-1\ne i\ne 2r}^p\sqrt{\th_i-\th_{2r}}}
\label{eqn:uni:onepair}
\end{equation}
and therefore there is no need to numerically integrate this term. This is not surprising as in this case the pole is of order 1 and therefore the residue theorem implies the result above.
Similarly, when $p$ is odd, the last term  corresponds to the branch cut $(-\infty, -\th_p)$ 
\begin{equation}
\intl^{\infty}_{\th_p}\frac{e^{-st}}{ \sqrt{-\prod_{i=1}^p(\th_i-t)} }  dt=\intl^{\infty}_{0}\frac{\text{exp}(-s\th_p-sv)}{ \sqrt{\prod_{i=1}^{p-1}(\th_i-\th_p-v) }\sqrt{v} }  dt=2 e^{-s\th_p} \intl_{0}^{+\infty} \frac{e^{-su^2} du}{\sqrt{\prod_{i=1}^{p-1}(\th_p-\th_i+u^2)}}
\label{eqn:uni:repar:inf}
\end{equation}
where and $t=\th_{p}+v$, $v=u^2$. 
These terms together with those of \eqref{eqn:uni:repar}, can also be efficiently integrated numerically as these functions are either decaying exponentially to zero or having a ``U" shape with only one critical point. See Figure~\ref{fig:g} for the log-scale behaviour of these elementary functions. 

	\subsection{Adding degrees of freedom (multiplicities of $\theta_1$) higher than 2 } 
A recursive application of the derivatives to the normalising constants as in \cite{KumeWood2007} or as generalised in \eqref{eqn:diff:property} could in principle generate the required expressions. 
The  differentiation needs to be applied to each of the elementary terms of Theorem~\ref{Theo:Bingh} and a simple differentiation has the effect adding the multiplicity by 2.
For example, the differentiation as below for each of the terms of Theorem~\ref{Theo:Bingh} leads to the cases of multiplicity 3: 
%
$$
\frac{\partial}{\partial \th_j} \intl_{C_r}  \frac{e^{t}}{ \sqrt{-\prod_{i=1}^p(\th_i+t)} }  dt
=\left \{
\begin{array}{lr}
-\frac{1}{2}\intl_{-\th_{2r}}^{-\th_{2r-1}}   \frac{1}{(\th_j+t)\sqrt{-\prod_{i=1}^p(\th_i+t)}}e^{st} dt & \th_{2r} \ne\th_ j\ne \th_{2r-1}\\ 
\frac{1}{\th_{2r}-\th_{2r-1}}\intl_{-\th_{2r}}^{-\th_{2r-1}} \frac{\th_{2r-1}+t}{\sqrt{-\prod_{i=1}^p(\th_i+t)}}e^{st}\left(-1+\frac{1}{2}  \sum\limits_{2r-1\ne i\ne 2r} \frac{1}{\th_i+t}\right)dt
 & j=\th_{2r-1}\\
 \frac{1}{\th_{2r}-\th_{2r-1}}\intl_{-\th_{2r}}^{-\th_{2r-1}} \frac{\th_{2r}+t}{\sqrt{-\prod_{i=1}^p(\th_i+t)}}e^{st}\left(-1+\frac{1}{2}  \sum\limits_{2r-1\ne i\ne 2r} \frac{1}{\th_i+t}\right)dt& j=\th_{2r}
\end{array} \right.
$$

\section{Cases of practical importance}

\subsection{Bingham distribution}
For the Bingham distributions with ordered and  distinct $\th_i$ we have that 
	\begin{eqnarray}
\mathcal{C}(\kbm{\th })
	&=& \intl_{\mathcal{S}^{p-1}} \text{exp}( \sum_{i=1}^p -\th_i x_i^2)d_{\mathcal{S}^{p-1}}(\x)=2 \pi^{\frac{\sum_{i=1}^{p} n_i}{2}-1}  \sum_{r=1}^{[\frac{p+1}{2}]}(-1)^{r+1} \int_{\theta_{2r-1}}^{\theta_{2r}}  \frac{e^{-t}}{\sqrt{-\prod_{i=1}^p (\th_i-t)}} dt.
\label{eqn:Bingh}
\end{eqnarray} 

If $p=3$, the Bingham on the ordinary sphere has normalising constant 
$$
\frac{\mathcal{C}(\kbm{\th })}{4 \sqrt{\pi} }
= e^{-\th_{1}}\int_{0}^{\pi/2}  \frac{\text{exp}(-(\th_2-\th_1)\sin^2u) du}{\sqrt{\th_3-\th_{1}\cos^2 u-\th_{2}\sin^2 u}} du- e^{-\th_3} \intl_{0}^{+\infty} \frac{e^{-u^2} du}{\sqrt{\prod_{i=1}^{2}(\th_3-\th_i+u^2)}}
$$
  
For $p=4$, there are only two finite branch cuts, and from \eqref{eqn:uni:repar} we can generate this expression as

\begin{eqnarray}
\frac{\mathcal{C}(\kbm{\th })}{4 \pi}&=&
 e^{-\th_{1}}\int_{0}^{\pi/2}  \frac{\text{exp}\left(-(\th_2-\th_1)\sin^2u\right) du}{\sqrt{\prodl_{i=3}^4(\th_i-\th_{1}\cos^2 u-\th_{2}\sin^2 u)}} du-e^{-\th_{3}}\int_{0}^{\pi/2}  \frac{\text{exp}\left(-(\th_4-\th_3)\sin^2u\right ) du}{\sqrt{\prodl_{i=1}^2(\th_i-\th_{3}\cos^2 u-\th_{4}\sin^2 u)}} du\\
\label{eqn:explicit:c4} 
&=&
\frac{ e^{-\th_{1}}}{2}
\int_0^1 \frac{e^{-u\alpha_1} du}{\sqrt{u(1-u)(\th_3-\th_{1}-u\alpha_1)(\th_4-\th_{1}-u\alpha_1)}}\\
&&
-\frac{e^{-\th_{3}}}{2} \int_0^1 \frac{e^{-u\alpha_2} du}{\sqrt{u(1-u)(\th_1-\th_{3}-u\alpha_2)(\th_2-\th_{3}-u\alpha_2)}}
\nonumber
\end{eqnarray}
where $\alpha_{1}=\theta_2-\theta_1$ and $\alpha_{2}=\theta_4-\theta_3$.
In the light of property \eqref{eqn:diff:property}, differentiating for various $\th_i$ here leads to other  representations corresponding to additional multiplicities. 

\subsection{Fisher distribution of rotations in $\R^3$ and its use in 3-D shape inference}
The expression of Bingham distributions for $p=4$, can be adopted as in \cite{Wood1993a} so that for 
some diagonal matrix $$\Phi=\left(\begin{array}{ccc}
\phi_1&0&0\\
0&\phi_2&0\\
0&0&\phi_3
\end{array}\right)\quad \text{and} \quad \kbm{\th}=-2(\phi_1, \phi_2,\phi_3,\phi_1+\phi_2+\phi_3)$$
the normalizing constant of the Fisher distribution in $SO(3)$ is 
$$
\int_{SO(3)} e^{\text{tr}(\Phi \kb{R})} d\kb{R}=\frac{1}{2}e^{\phi_1+\phi_2+\phi_3}\mathcal{C}(\kbm{\th })\quad 
$$
where the integration is taking place along the Haar uniform measure $d\kb{R}$ on the group of rotations $ SO(3)$ and the evaluation of $\mathcal{C}(\kbm{\th })$ is as in \eqref{eqn:explicit:c4}. This formula is similar to the one shown in \cite{Wood1993a}.  

In \cite{Drydenb}, the  likelihood function for a very general set of regression shape models is based on the efficient evaluation of this normalising constant. 
In addition they address the bias correction of the well known procrustes mean shape estimators for the 3-d shape models based on the principles of the maximum likelihood estimation. 
For establishing the resulting estimators the authors suggest using the gradient of $\log \int_{SO(3)} e^{\text{tr}(\Phi R)} dR$ as a bias correcting term. In particular they use this result 
\[
\nabla_{\Phi} \log \int_{SO(3)} e^{\text{tr}(\Phi \kb{R})} d\kb{R}=
\left(\begin{matrix}
1-\frac{2}{\mathcal{C}(\kbm{\th })} \frac{\partial \mathcal{C}(\kbm{\th })}{\partial \theta_1}-\frac{2}{\mathcal{C}(\kbm{\th })}\frac{\partial \mathcal{C}(\kbm{\th })}{\partial \theta_4}\\
1-\frac{2}{\mathcal{C}(\kbm{\th })} \frac{\partial \mathcal{C}(\kbm{\th )}}{\partial \theta_2}-\frac{2}{\mathcal{C}(\kbm{\th })}\frac{\partial \mathcal{C}(\kbm{\th })}{\partial \theta_4}\\
1-\frac{2}{\mathcal{C}(\kbm{\th })} \frac{\partial \mathcal{C}(\kbm{\th )}}{\partial \theta_3}-\frac{2}{\mathcal{C}(\kbm{\th })}\frac{\partial \mathcal{C}(\kbm{\th })}{\partial \theta_4}
\end{matrix})
\right)
\]
where the partial derivative expressions are easily obtained from the integral representations as in \eqref{eqn:Bingh}.  
In fact  \cite{Drydenb} show that that for practical purposes,  these components are accurately  and very easily calculated using the saddlepoint approximation. 

%
%
%

\subsection{Kent's formula for Complex Bingham }
If even multiplicities are introduced, the limit of the expression \eqref{eqn:Bingh} will suffice.  
For example, 
if all $\th_{i}$ as in Lemma 1 are of even multiplicities then the corresponding function $g(t)$ will not have any branch cuts but just simple poles therefore a simple application of the residue theorem generates the wellknown result of Kent for the complex Bingham distributions. More specifically, if $p=2k$ while $\th_i$'s are coalescing in pairs i.e. $\th_{2r}\to \th_{2r-1}$, and following similar arguments to those in \eqref{eqn:uni:onepair}, each term of \eqref{eqn:Bingh}  becomes
$$
\frac{e^{-\th_{2r}}}{\prodl_{2r-1\ne i\ne 2r}^p(\th_i-\th_{2r})^{1/2}}
\int_0^1 \frac{du}{\sqrt{u(1-u)}}=\frac{e^{-\th_{2r}}}{\prodl_{ 2r-1\ne i\ne 2r}^k(\th_i-\th_{2r})} \pi
$$
and therefore for $k$ distinct pairwise $\theta$'s.
\begin{equation}
\Cth=2 \pi^{k} \sum_{r=1}^k (-1)^{r+1} \frac{e^{-\th_{r}}}{\prodl_{ i\ne r}^p(\th_i-\th_{r})} =2 \pi^{k} \sum_{r=1}^k  \frac{e^{-\th_{r}}}{\prodl_{ i\ne r}^p(\th_r-\th_{i})}
\label{eqn:complex:Bingh}
\end{equation}
which is consistent with the result of Kent for the complex Bingham distribution.  Note that in the light of \eqref{eqn:dens:sph_nc} for $n_i=2$ and $\gamma_i=0$,   Kent's normalising constant above can also be used for close form expressions of the pdf  and cdf of  a positive linear combination of the exponential distributing terms see e.g. \cite{Kent:94}.  
\subsection{Kent distributions}
In these distributions, we have only one $\g_{r}\ne0$ and in  particular for the $p=3$ case,  $\g_{2}\ne 0$ and therefore, the integration along the unbounded branch cut is reduced as in Theorem 2. 

An additional feature for the Kent distribution is that the middle pole $\th_2$ is equally distant from the other two so that $\kbm{\th}=(\th_1,\th_1+\alpha, \th_1+2\alpha)$. Since the non-central term is not on the unbounded branch $(-\infty, -\th_3)$, one can reduce that integration along the real line throughout this interval and use an additional circular contour parametrisation $t=-\th_1-\frac{\alpha}{2}+re^{\textbf{i}u}$ for the finite branch, and so for the Kent distribution:

 \begin{eqnarray*}
 	\frac{\mathcal{C}(\kbm{\th }, \gamma_2)}{2 e^{-\th_1} \sqrt{\pi}}
 	&=& \frac{-r e^{-\frac{\alpha}{2}} }{2 \pi}  \intl_{0}^{2\pi}  \textbf{Im}\left( \frac{\text{exp}\left(\frac{\g_{2}^2}{4(\frac{\alpha}{2}+re^{\textbf{i}u})}+re^{\textbf{i}u}\right )}{\sqrt{(-\frac{\alpha}{2}+re^{\textbf{i}u})(\frac{\alpha}{2}+re^{\textbf{i}u})(3\frac{\alpha}{2}+re^{\textbf{i}u})}}\right)  du\\
 	&-&  2e^{-2\alpha} \intl_{0}^{+\infty} \frac{\text{exp}\left(-u^2-\frac{\g_{2}^2}{4(\alpha+u^2)}\right )du}{\sqrt{(\alpha+u^2)(2\alpha+u^2)}}
 \end{eqnarray*}
where the radius $\frac{\alpha}{2}<r<\alpha$ is chosen so that the circle contains only the first branch cut. 
Note that for implementing this distribution in directional statistics, $\th_{1}$ can be allowed to  be $0$ and therefore the normalizing constant of this distribution is in fact defined in terms of only two parameters $\alpha$ and $\gamma_{2}$. These parameters correspond to $\beta$ and $\kappa$ in \cite{K:82}. 


\def \GenExp{1}
\if\GenExp0
	\subsection{Some related expectations for the central case}
\AKL{The following part in blue might not necessarily fully link to the asymptotic testing yet}
Suppose that the interest is to evaluate expectations of the type
	\[
	\int_{a}^{b} f(s) \mathcal{C}(\xi s \kbm{\lambda}) ds
	\] 
for some positive valued function $f(s)$ where typically $a=0$ and $b=\infty$. 
A general expression follows from
	\[
	\mathcal{C}(s \kbm{\lambda})=s^{1-p/2} \sum_{r=1}^{[\frac{p+1}{2}]}(-1)^{r+1} \int_{\lambda_{2r-1}}^{\lambda_{2r}}  \frac{e^{-\xi st}}{\sqrt{-\prod_{i=1}^p (\lambda_i-t)}} dt
	\]
	so that 
	\[
	\int_{0}^{\infty} f(s) \mathcal{C}(s \kbm{\lambda}) ds=
	\int_{0}^{\infty} f(s) s^{1-p/2} \sum_{r=1}^{[\frac{p+1}{2}]}(-1)^{r+1} \int_{\lambda_{2r-1}}^{\lambda_{2r}}  \frac{e^{-\xi st}}{\sqrt{-\prod_{i=1}^p (\lambda_i-t)}} dt ds
	\]
Hence,
	\[\int_{0}^{\infty} f(s) \mathcal{C}(s \kbm{\lambda}) ds=
	\sum_{r=1}^{[\frac{p+1}{2}]}(-1)^{r+1} \int_{\lambda_{2r-1}}^{\lambda_{2r}}   \frac{g(t)}{\sqrt{-\prod_{i=1}^p (\lambda_i-t)}} dt
	\]
	where $g(t)=\int_{0}^{\infty} s^{1-p/2} f(s) e^{-\xi st} ds$.	
If for example 	$f(s)=s^\alpha e^{-\beta s}$  then 
$$
g(t)=\int_{0}^{\infty} s^{\alpha-p/2+1} e^{-\beta s} e^{-\xi  st} ds= (\beta+\xi t)^{p/2-\alpha-2} \int_{0}^{\infty} y^{\alpha-p/2+1} e^{-y}dy=(\beta+ \xi t)^{p/2-\alpha-2} \Gamma(\alpha-p/2+2)
$$
takes a closed  form as in the expression for $f(\kbm{\mu}, h)$ (see in the following section) and if 
$f(s)=s^\alpha e^{-\beta  s^2/2}$ then from 3.462 in \cite{AS}
\[
g(t)=\int_{0}^{\infty} s^{\alpha-p/2+1} e^{-\beta  s^2/2} e^{-st} ds= \beta^{\frac{\alpha-p/2+2}{2}}\Gamma(\alpha-p/2+2)e^{\frac{t^2}{4\beta}}D_{-\alpha+p/2-2}(\frac{-t}{\sqrt{\beta}})
 \]
where $D_{-\alpha+p/2-2}$ is the parabolic cylinder function.
A similar expression is in fact related to the likelihood for the shape integral for 3 dimensional objects where $p=4$ and $\alpha=3k/2-1$ where $k+1$ is the number of landmarks. 

In the following we will focus on two particular situations; one with applications is the sphericity testing and the other one for a special family of distributions on the complex projective spaces. 

\subsection{Asymptotic sphericity testing}

 Consider a multivariate distributed vector $N_p(0,\sigma^2(I_p+hvv'))$, where $\sigma, h>0$ and $v$ is some vector of norm 1. 
 The focus here is to test whether $H_0: h=0$ against $H_1:h>0$, which is equivalent to  spherical symmetry of the distribution with the corresponding sample covariance matrix. This is studied in  \cite{Onatski2013} where the sample size $n$ and dimensionality $p$ are both reaching infinity at some controllable relative rate. Let now $X$ be some $p\times n$ random matrix with i.i.d. Gaussian columns $N_p(0,\sigma^2(I_p+hvv'))$. Let also $\lambda_1\ge\lambda_2\ge\cdots \ge\lambda_p$ be the ordered eigenvalues of $\frac{XX^t}{n}$ with vector form $\kbm{\lambda}=(\lambda_1,\lambda_2, \cdots\lambda_p)$ and $\kbm{\mu}=(\mu_1,\mu_2, \cdots\mu_p)$ where $\mu_i=\frac{\lambda_i}{\sum_{i=1}^{p}\lambda_i}$. Based on the Proposition 1 there,  the corresponding density functions for $\kbm{\lambda}, h$ and $\kbm{\mu},h$ are
\begin{eqnarray*}
f(\kbm{\lambda}, h)&\propto& \frac{1}{(1+h)^{n/2}}\int_{S^{p-1}}e^{\xi v'\kbm{\lambda}v}dv=\frac{1}{(1+h)^{n/2}}\mathcal{C}(-\xi\kbm{\lambda})=e^{\xi C}\mathcal{C}(\xi C-\xi\kbm{\lambda})\quad  \forall C\ge \lambda_1
\end{eqnarray*}
\AK{
and
\[
	f(\kbm{\mu}, h)\propto \frac{1}{(1+h)^{n/2}}\int_0^\infty y^{np/2-1}e^{-ny/2}\int_{S^{p-1}}e^{y\xi v' \kbm{\mu}v}dv dy
\]
with $\xi=\frac{n}{2}\frac{h}{1+h}$.  The likelihood ratio statistics for the relevant hypothesis test are simply $\frac{f(\kbm{\lambda}, h)}{f(\kbm{\lambda}, 0)}$ and $\frac{f(\kbm{\mu}, h)}{f(\kbm{\mu}, 0)}$, which by using the contour integration are derived as
\begin{result}
\begin{equation*}
	\frac{f(\kbm{\lambda}, h)}{f(\kbm{\lambda}, 0)}=  \frac{\xi^{1-\frac{p}{2}}} {(1+h)^{n/2}}\sum_{r=1}^{[\frac{p+1}{2}]}(-1)^{r+1} \int_{\lambda_{2r}}^{\lambda_{2r-1}}  \frac{e^{t\xi}}{\sqrt{-\prod_{i=1}^p (t-\lambda_i)}} dt\quad \xi=\frac{n}{2}\frac{h}{1+h}
\end{equation*}
and
\begin{equation}
	\frac{	f(\kbm{\mu}, h)}{	f(\kbm{\mu}, 0)}=\frac{E(X_h^{n(p-1)/2})}{\prod_{i=1}^{p}\th_i^{\frac{1}{2}}(1+h)^{n/2} }=\frac{ \sum_{r=1}^{[\frac{p+1}{2}]}(-1)^{r+1} \int_{\theta_{2r-1}}^{\theta_{2r}}  \frac{1}{t^{np/2} \sqrt{-\prod_{i=1}^p (\th_i-t)}} dt}{(1+h)^{n/2}}
\end{equation}
and $\kbm{\th}=\kb{1}-\frac{h}{1+h}\kbm{\mu}$ and $X=Y^tY$ where $Y\sim N_p(\kb{0},\frac{1}{2\kbm{\th}})$. 
\end{result}
}
\proof -to be moved to Appendix

Note that  by rescaling the integrating variable in \eqref{eqn:Bingh} by some constant $\xi$ the branch cuts are determined by increasing $-\lambda_i$ and  accordingly rescaled as 
\begin{eqnarray*}
	f(\kbm{\lambda}, h)\propto \frac{\xi^{1-p/2}}{(1+h)^{n/2}} \sum_{r=1}^{[\frac{p+1}{2}]}(-1)^{r+1} \int_{-\lambda_{2r-1}}^{-\lambda_{2r}}  \frac{e^{-t\xi}}{\sqrt{-\prod_{i=1}^p (-\lambda_i-t)}} dt\\
	\propto \frac{\xi^{1-p/2}}{(1+h)^{n/2}} \sum_{r=1}^{[\frac{p+1}{2}]}(-1)^{r+1} \int_{\lambda_{2r}}^{\lambda_{2r-1}}  \frac{e^{t\xi}}{\sqrt{-\prod_{i=1}^p (t-\lambda_i)}} dt
\end{eqnarray*}

\begin{eqnarray*}
	f(\kbm{\mu}, h)&\propto& \frac{1}{(1+h)^{n/2}}\int_0^\infty y^{np/2-1}e^{-ny/2}\int_{S^{p-1}}e^{y\xi v' \kbm{\mu}v}dv dy=\frac{1}{(1+h)^{n/2}} \int_0^\infty y^{np/2-1}\int_{S^{p-1}}e^{-y\frac{n}{2}v'(\kb{1}-\frac{2}{n} \xi  \kbm{\mu})v}dv dy\\
	&=&
	 \frac{1}{(1+h)^{n/2}}\int_0^\infty  y^{np/2-1} \mathcal{C}(y \frac{n}{2}\kb{\th}) dy=\frac{1}{(1+h)^{n/2}}\int_0^\infty  (\frac{y}{n/2})^{np/2-1} \mathcal{C}(y \kb{\th}) d(\frac{y}{n/2})\\
	 &=&\frac{(n/2)^{-np/2}}{(1+h)^{n/2}}\int_0^\infty  y^{np/2-1} \mathcal{C}(y \kb{\th}) d(y)=\frac{(n/2)^{-np/2}}{\prod_{i=1}^{p}\th_i^{\frac{1}{2}}(1+h)^{n/2}}\int_0^\infty  y^{np/2-1} y^{1-p/2} f_{\kb{\th}}(y) dy\\
	 &=&\frac{(n/2)^{-np/2}}{\prod_{i=1}^{p}\th_i^{\frac{1}{2}}(1+h)^{n/2}} \int_0^\infty  y^{n(p-1)/2} f_{\kb{\th}}(y) dy=\frac{(n/2)^{-np/2}}{\prod_{i=1}^{p}\th_i^{\frac{1}{2}}(1+h)^{n/2}}E(X^{n(p-1)/2})
\end{eqnarray*}
$\xi=\frac{n}{2}\frac{h}{1+h}$ and $\kbm{\th}=\kb{1}-\frac{h}{1+h}\kbm{\mu}$ and $X$ as in \eqref{eqn:gen:rep}, namely, $X=Y^tY$ where $Y\sim N_p(\kb{0},\frac{1}{2\kbm{\th}})$.  
\AK{The integral representation is obtained as 
\begin{eqnarray*}
	f(\kbm{\mu}, h)
	&\propto&\frac{(n/2)^{-np/2}}{(1+h)^{n/2}}\int_0^\infty  y^{np/2-1} \mathcal{C}(y \kb{\th})dy
	\\
	 &\propto& \frac{(n/2)^{-np/2}}{(1+h)^{n/2}} \sum_{r=1}^{[\frac{p+1}{2}]}(-1)^{r+1} \int_{0}^{\infty} \int_{\theta_{2r-1}}^{\theta_{2r}}  y^{np/2-1} \frac{e^{-ty}}{\sqrt{-\prod_{i=1}^p (\th_i-t)}} dt	dy	\\	&=& \frac{(n/2)^{-np/2}\Gamma(np/2)}{(1+h)^{n/2}} \sum_{r=1}^{[\frac{p+1}{2}]}(-1)^{r+1} \int_{\theta_{2r-1}}^{\theta_{2r}}  \frac{1}{t^{np/2} \sqrt{-\prod_{i=1}^p (\th_i-t)}} dt	
\end{eqnarray*}
with the infinite branch cut if $p$ odd stretching to the $\infty$ value. 
Note that modulo the spherical volume, at $h=0$,  $f(\kbm{\lambda}, 0)=1$ and $f(\kbm{\mu}, 0)=\Gamma(np/2)$ and the ratios of the result follow.
}
\endproof

%
%
 
 These results indicate that the corresponding contour integration used in Lemmas 1,2,3 and 4 in \cite{Onatski2013} depend on only $\kbm{\lambda}$ and  not on the parameter $h$ and therefore the asymptotic analysis can be carried out without adopting any integration contour depending on $h$. 
 This fact is used in Lemma 4 of that paper which focuses on the asymptotic behaviour of $\frac{f(\kbm{\lambda}, h)}{f(\kbm{\lambda}, 0)}$ and $\frac{f(\kbm{\mu}, h)}{f(\kbm{\mu}, 0)}$ as both $p$ and $n$ go to $\infty$ such that $\frac{p}{n}\to c$. \AK{Note the latter ratio reduces to evaluating the $np/2$ moment for the linear combination of chi-squares determined by $\kbm{\th}$  for $h>0$. Such ratios for integer values of $np/2$ are available in closed form and the asymptotic results can be easily derived or improved. Similarly, one can easily obtain the relevant p-values for the test statistics by using a straightforward simulation procedure.}
\AKL{Could we improve on the limit of this ratio as $n\to \infty$ and $p\simeq c n$?		
Andy, I seem to remember that you suggested  to consider numerically this limit for large n and p? The code in R does not seem to reliably work for p a few hundreds, but it could be just a matter of reparametrisation or we could use SPA?
}

\fi

\def \ComQuart{1}
\if\ComQuart 0

\subsubsection{Complex Quartic distribution}
Such distributions are defined for the modelling of data in the shape distributions of planar configurations with finite number of landmarks, $p$ say.
\[
f(X)=c_{CBQ}(\Omega)^{-1}\exp \{ -\frac{1}{2}\left( X^t\Omega X-(X^t X)X^T\Omega^{(as)}X \right) \}
\]
The  matrix $\Omega$ is a $ 2(p-1)\times2(p-1) $-dimensional real symmetric matrix, 
\[ 
c_{CBQ}(\Omega)=\frac{1}{2}\int_0^1 s^{p-2} \mathcal{C}(\kbm{\th}(s))ds
\]
where $\kbm{\th}(s)$ are the eigenvalues of $\Psi(s)=-\frac{1}{2}(s\Omega-s^2\Omega^{(as)})$ and 
\[
\Omega^{(as)}=\frac{1}{2}\left(
\begin{array}{cc}
\Omega_{11}-\Omega_{22}&\Omega_{12}+\Omega_{21}\\
\Omega_{21}+\Omega_{12} & \Omega_{22}-\Omega_{11}
\end{array}
\right)=\frac{1}{2}\Omega-\frac{1}{2}\mathcal{I}
\Omega
\mathcal{I} \quad \mathcal{I}=\left(\begin{array}{cc}
0&\kbm{I}_{p-1}\\
\kbm{I}_{p-1} & 0
\end{array}
\right)
\]
From the definitions 
\[
\mathcal{C}(\kbm{\th}(s))=\int_{S^{p-2}} \exp\{-tr(X(s\Omega-\frac{s^2}{2} (\Omega-\mathcal{I}\Omega \mathcal{I}^T))X^T)\}dX
\]
Similar to the method mentioned in ??? where the Saddle point approximation was implemented, for chosen grid of points $0<s_1<s_2...<s_n<1$ a two step numerical integration procedure can be easily adopted as
$$
\hat{c}_{CBQ}(\Omega)=\frac{1}{2} \sum_{i=1}^{n} s_i^{p-2} \mathcal{C}(\kbm{\th}(s_i))
$$
where $\mathcal{C}(\kbm{\th}(s_i))$ are evaluated numerically using the standard numerical integration routines.

\fi

\def \showdiff{1}
\if\showdiff1

\section{Difference of two positive linear combinations}
	If in general some of the $\lambda_i$ in \eqref{eqn:gen:rep} are negative then the characterisation of the problem in terms of normal distribution components is as the difference between two quadratic forms.   The contour integration however  can be defined along the vertical lines such that the strictly negative poles are on the left of the contours $\textbf{i}\R+t_0$  while $t_0<0$ as in \eqref{eqn:dens}. More specifically,
	if $Z=X-Y$ where both $X$ and $Y$ have the same general expression as in \eqref{eqn:gen:rep} for some positive parameters $\kbm{\th}$, $\kbm{\g}$, $\kbm{\th}'$ and $\kbm{\g}'$, with respective densities
	\begin{equation}
		f_{\mybth, \mybg^2,\kbm{n}}(s)= \kappa
		\frac{1}{ 2\pi \textbf{i}}\intl_{\textbf{i}\R+t_0}  g(t)   e ^{st} dt \quad f_{\mybth', \mybg'^2,\kbm{n}'}(s)=\kappa'  \frac{1}{ 2\pi \textbf{i}} \intl_{\textbf{i}\R+v_0 }  g'(v)  e ^{sv} dv \quad s>0
		\label{eqn:dens1and2}
	\end{equation}
	where $\kappa=\prod_{i=1}^{p} \th_i^{\frac{n_i}{2}} \text{exp}(-\sum_{i=1}^{p}\frac{\gamma^2_i}{4\th_i}) $, $\kappa'=\prod_{i=1}^{p'} \th_i'^{\frac{n'_i}{2}}\text{exp}(-\sum_{i=1}^{p}\frac{\gamma'^2_i }{4\th'_i }) $ and 
	$$
	g(t)=\frac{\text{exp}(\sum_{i=1}^p \frac{\gamma_i^2}{4(\th_i+t)})}{\prod_{i=1}^p(\th_i+t)^{\frac{n_i}{2}} }
	\quad g'(v)=\frac{\text{exp}(\sum_{i=1}^p \frac{\gamma'^2_i}{4(\th'_i+v)})}{\prod_{i=1}^p(\th'_i+v)^{\frac{n_i'}{2}} } $$
	the following result holds:
	\begin{theorem}
		If  $X=\sum_{i=1}^{p}
		\frac{1}{2(\theta_i)} \chi^2_{n_i}(\frac{\gamma_i^2}{2 (\th_i)})$ and  $Y=\sum_{i=1}^{p'}
		\frac{1}{2(\theta'_i)} \chi^2_{n'_i}(\frac{\gamma_i'^2}{2 (\th'_i)})$
		then the pdf  of $Z=X-Y$ at $z\ge 0$ is 
		$$
		f_{Z}(z)=\prod_{i=1}^{p} \theta_i^{\frac{n_i}{2}} \exp(-\sum_{i=1}^{p}\frac{\gamma^2_i}{4\th_i}) \prod_{i=1}^{p} \th_i'^{\frac{n'_i}{2}}\exp(-\sum_{i=1}^{p}\frac{\gamma'^2_i }{4\th'_i }) \frac{-1}{2 \pi\textbf{i}} \intl_{\Gamma}   g(t) g'(-t)e^{zt}  dt $$
		where 
		the integrating contour $\Gamma$ can be simplified as in Theorems 1 and 2 along branch cuts involving only the values of $\kbm{\th}$.
		Additionally
		$$
		P(X-Y>z\ge0)=\frac{e^{\th_0 z}}{\th_0}f_{\tilde{Z}}(z)\prod_{i=1}^p\frac{\sqrt{\th_i}}{\sqrt{\th_i+\th_0}} \prod_{i=1}^{p'}\frac{\sqrt{\th'_i}}{\sqrt{\th'_i-\th_0}}\exp(\sum_{i=1}^{p}\frac{\gamma^2_i}{4\th_i}-\frac{\gamma^2_i}{4(\th_i+\th_0)} +\sum_{i=1}^{p'}\frac{\gamma'^2_i }{4\th'_i }-\frac{\gamma'^2_i }{4(\th'_i-\th_0)} )
		$$
		for any arbitrary $0<\th_0<min(\kbm{\th}')$ and $$\tilde{Z}=\frac{1}{2\theta_0} \chi^2_2(0)+\sum_{i=1}^{p}
		\frac{1}{2(\theta_i+\theta_0)} \chi^2_{n_i}(\frac{\gamma_i^2}{2 (\th_i+\th_0)})-\sum_{i=1}^{p'}
		\frac{1}{2(\theta'_i-\theta_0)} \chi^2_{n'_i}(\frac{\gamma_i'^2}{2 (\th'_i-\th_0)}).$$
		The corresponding expressions for $z<0$ can be obtained similarly for  $f_{Y-X}(-z)$ by switching the roles of $X$ and $Y$ as in $-Z=Y-X$. 
		\label{Theorem:diff}
	\end{theorem}
	As a special case one could think of the exponential distributions which correspond modulo some scaling to a central chi-square of degree.  These cases have practical implications due to their connections with the Erlang distributions and Phase Type distributions in the probability theory. 
A recent paper for generating the distribution for the positive entries as in \cite{Neumueller2022} and also the proof for the extended case of real coefficients can be found in~\cite{Mathai1983}. 

In our construction however,  these expressions are simply the residues of ordinary poles.	Similar to the case of complex Bingham distributions which are also related to the exponentially distributed components,  the respective branch cuts reduce to simple poles and  the residues generate closed form expressions:
	\begin{corollary}
		If the chi-square components  in Theorem \ref{Theorem:diff} are central chi-squares r.v's  with degrees of freedom  2, namely $ Z=X-Y $  represents a linear combination of exponential random variables, then  the corresponding density and distribution functions at $z\ge0$  are evaluated as 
		$$
		f_{Z}(z)
		=\prod_{i=1}^{p} \theta_i  \prod_{j=1}^{p'} \th_j' 
		\sum_{i=1}^{p}(-1)^i\frac{e^{-\th_i z}}{\prod_{j\ne i}^{p}(\th_{j}-\th_i)\prod_{k=1}^{p'}(\th'_{k}-\th_i)}
		$$
		and
		$$
		P(Z>z \ge 0)
		=\prod_{i=1}^{p} \theta_i  \prod_{j=1}^{p'} \th_j' 
		\sum_{i=1}^{p}(-1)^i\frac{e^{-\th_i z}}{\th_i  \prod_{j\ne i}^{p}(\th_{j}-\th_i)\prod_{k=1}^{p'}(\th'_{k}-\th_i)}.
		$$
		For the cases when $z<0$ the roles of $\kbm{\th}$ and $\kbm{\theta'}$ are reversed.
	\end{corollary}
\fi

\section{Numerical evidence}

\begin{figure}[th!] 
	\centering
	\begin{minipage}{.5\textwidth}
	\centering
	\includegraphics[width=0.52\linewidth]{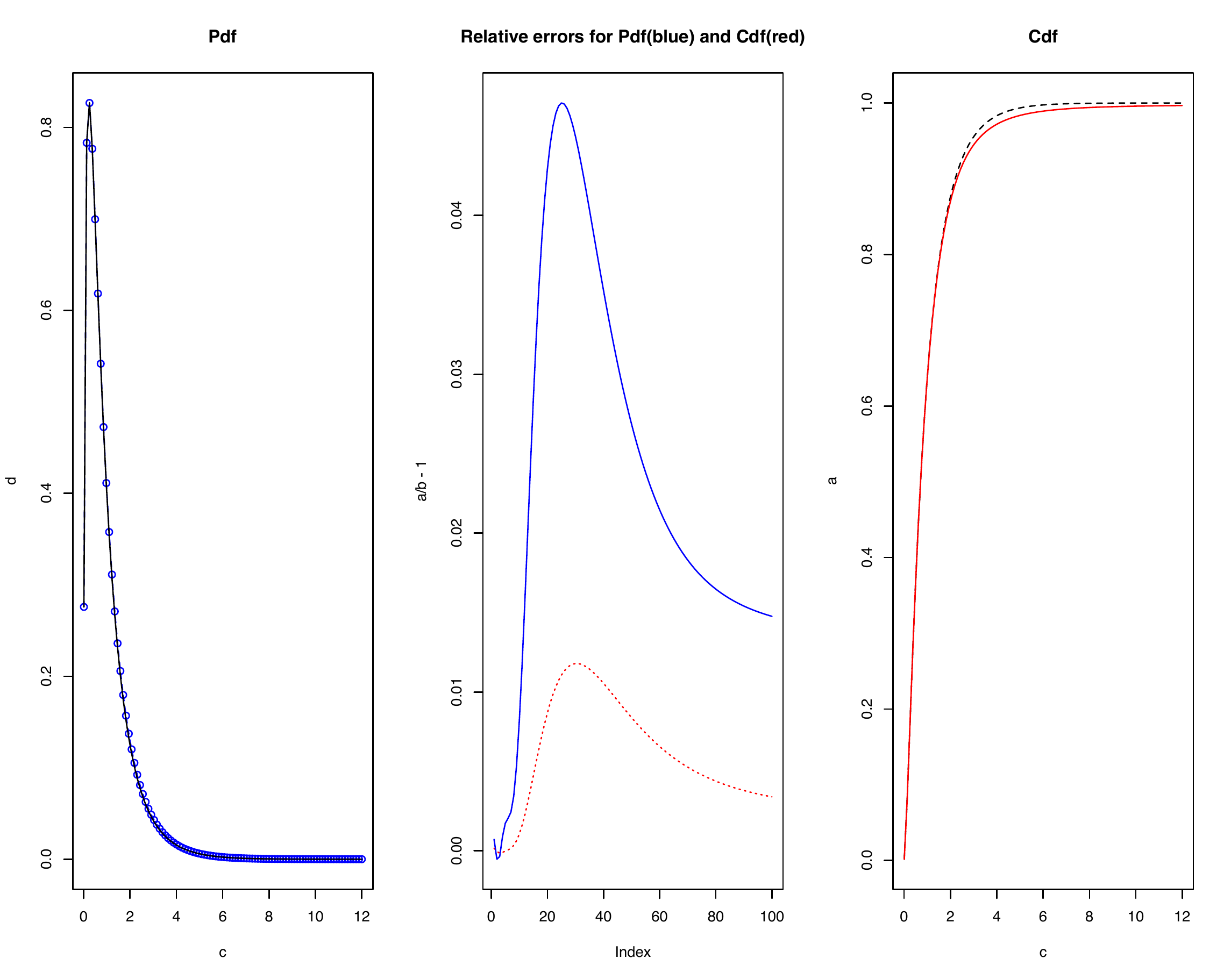}
	\caption{Comparisons for $\mybth_1$}
	\label{fig:plotq1}
\end{minipage}%
\begin{minipage}{.5\textwidth}
	\centering
	\includegraphics[width=0.52\linewidth]{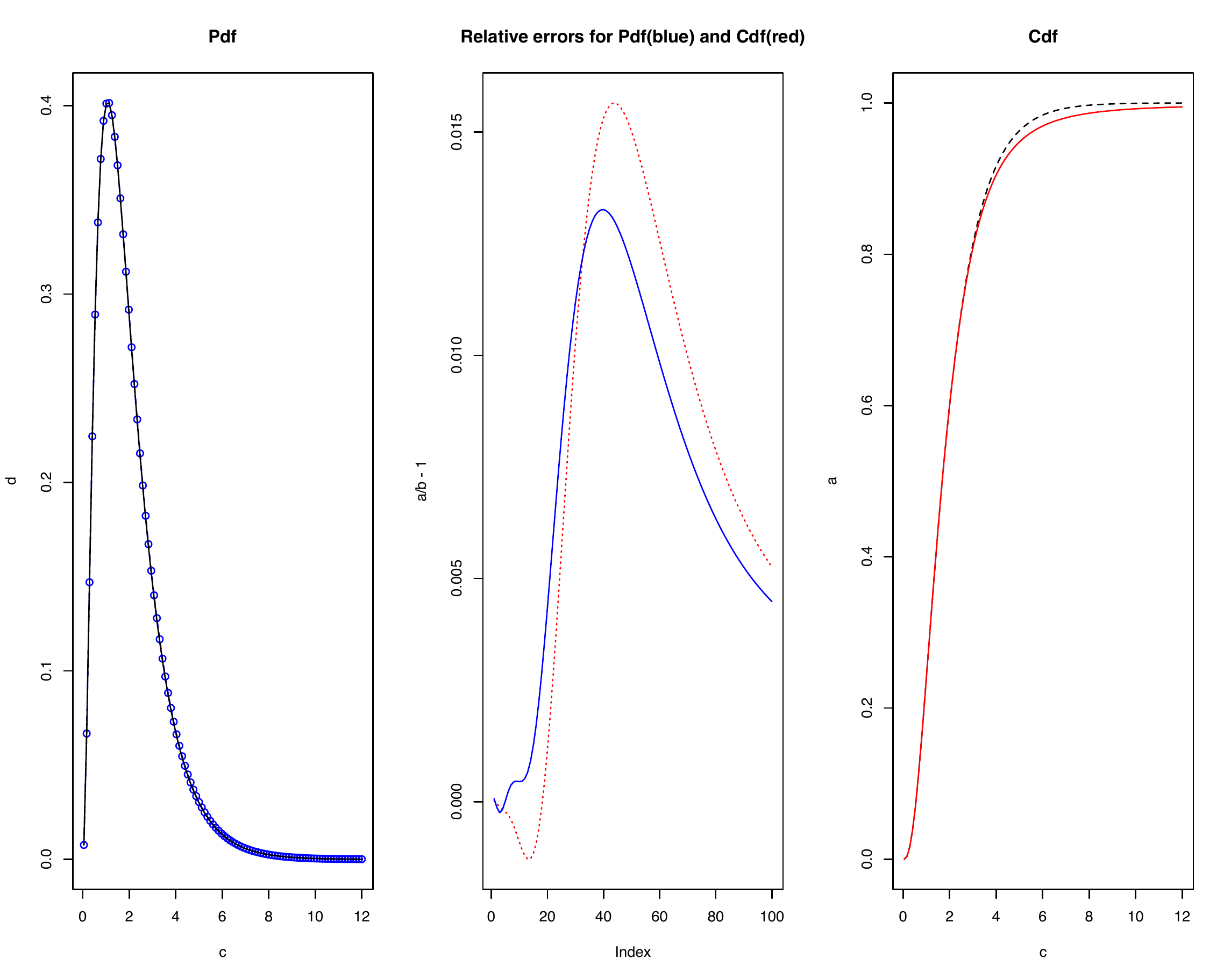}
	\caption{Comparisons for $\mybth_2$}
	\label{fig:plotq2}
\end{minipage}
\end{figure}

We illustrate our method by replicating the numerical figures reported in \cite{IMH:61}. We focus initially on the first two central cases reported there, where the coefficients lambda of $\eqref{eqn:gen:rep}$ are defined as $\kb{\lambda}_i=\frac{1}{2 \mybth_i}$ which in our parametrization they correspond to:
$$\mybth_1=(0.8333333,1.6666667,5.0000000) \quad \mybth_2=(0.8333333 ,0.8333333 ,1.6666667 ,1.6666667 ,5,5)$$
Note in particular that for $\kbm{\th}_2$ above,  the three distinct values have multiplicity $2$ i.e. simple poles and therefore as we have noted in the corresponding inversion is immediately available in closed form for both pdf and cdf using \eqref{eqn:complex:Bingh} and \eqref{eqn:dens:sph_nc}.  
In \cite{IMH:61} only the cdf values for specific entries are shown while we can see that the whole function can be easily obtained using our simple contour integrating terms.  
The comparison of both exact and SPA methods for density, pdf. and cdf. together with the corresponding  relative error are shown in Figures \eqref{fig:plotq1} and \eqref{fig:plotq2}. It can be easily seen that the density SPA used for the cdf calculations as stated in our Theorem~1, performs very well especially in the extreme values of the variable $s$. 

\def\myknitr{1}
\if\myknitr1

%
%
%
%

%

As illustrative evidence for the difference of two positive linear combinations, we can mention cases reported in \cite{IMH:61} where in particular, the three example entries on our Table~\ref{tab:table1} here relate to the cases $\frac{Q_3}{3}-\frac{2 Q_4}{3}$ and $\frac{Q_5}{2}-\frac{ Q_6}{2}$ and $\frac{1}{6}Q_3+\frac{2}{6}Q_6-\frac{1}{6}Q_5-\frac{2}{6}Q_4$ there. 


%
The approximation method reported in ~\cite{IMH:61} is illustrated against some exact figures which are also reported in our Table~\ref{tab:table1}. These exact values which the Imhof's method is comparing to are in full agreement with our method of bounded contour integration as seen for various values $s$. Our method can easily produce the pdf and cdf of these three cases as shown on Figure~8.

\begin{table}[ht!]
	\begin{center}
		\caption{Some numerical checks with those of \cite{IMH:61} }
		\label{tab:table1}
		\scalebox{0.8}{
		\begin{tabular}{l|c|c|r|r} 
			\text{Model}&\text{Our parametrisation  }$\mybth^{\kb{n}}_\kb{\gamma}$& s & \text{\cite{IMH:61}}&\text{Our Method}\\
			\hline
			$ \frac{1}{3}Q_3-\frac{2 }{3}Q_4 $ &$ 0.4^2_0+ 0.2^4_0+(\frac{2}{30})^6_0 -0.35^1_6 -0.15^1_2 $& -2 &0.9102 &0.9102254\\
			&& 0&0.4061 & 0.4061061\\
			&& 2.5 & 0.097&0.0097598\\
			\hline
			$ 	\frac{Q_5}{2}-\frac{ Q_6}{2} $&$ 3.5^6_6+ 0.15^2_2 -0.35^1_6-0.15^1_2 $& -2 & 0.9218&0.921792\\
			&& 2 & 0.4779&0.4778933\\
			&& 7 & 0.0396&0.0396319\\		
			\hline
			$ \frac{1}{6}(Q_3-Q_5)+\frac{2}{6}(Q_6-Q_4) $&$ 0.1_2^7+ 0.05^4_0+(\frac{0.1}{6})^2_0+ (\frac{1.4}{6})^1_6 $& -3 & 0.9861&0.9861469\\
			&$-0.2_0^2-0.1^4_0-(\frac{0.2}{6})^6_0-(\frac{0.7}{6})^6_6-0.05^2_2$& 0 & 0.5170&0.5170232\\
			&& 4 & 0.0152&0.0152041\\
			
		\end{tabular}}
	\end{center}
\end{table} 
%
%

\begin{figure}
	\centering
\includegraphics[width=0.6\textwidth]{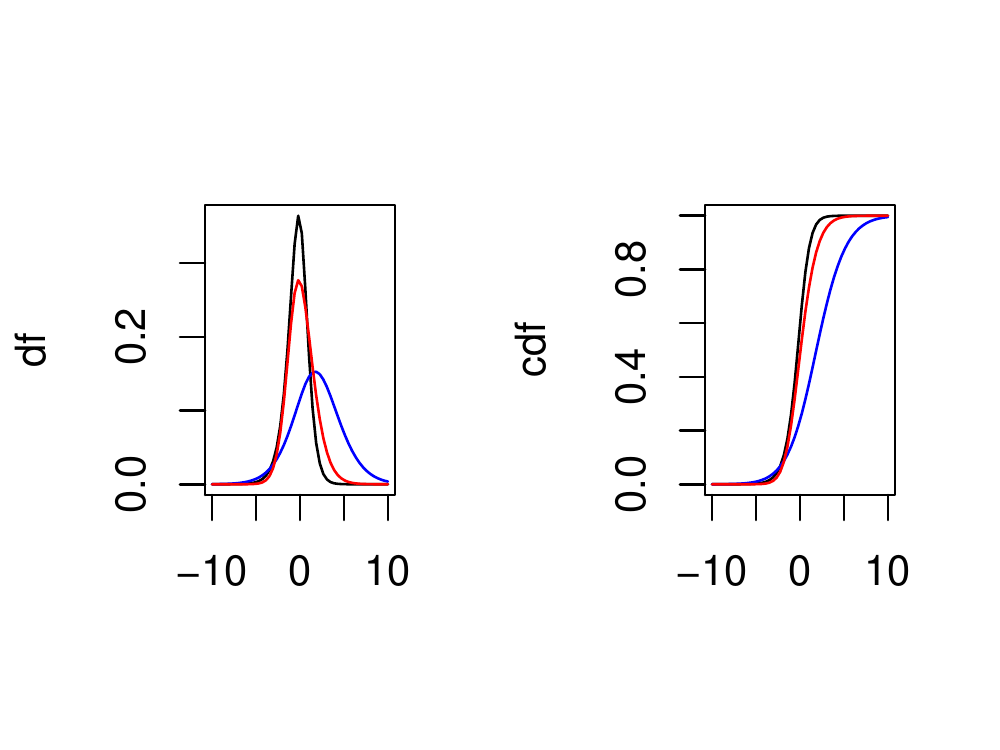}
	\label{figdif_qd}
	\caption{ A plot of pdf and cdf for$ \frac{1}{3}Q_3-\frac{2 }{3}Q_4 $ in black; $\frac{Q_5}{2}-\frac{ Q_6}{2} $ in blue and  	$ \frac{1}{6}(Q_3-Q_5)+\frac{2}{6}(Q_6-Q_4) $ in red.}
\end{figure}

\fi

\def \GenNotes{1}
\if\GenExp0

\section{Some notes on inversion process-Not to be included}
\large{The numerical inversion for general cases can be obtained while noticing that for $t=a+\text{i}b$
$$
Arg(\frac{1}{(\th_i+t)^{n_i/2}})=-\frac{n_i}{2}Arg(\th_i+t)=-\frac{n_i}{2}\tan^{-1}(\frac{b}{\th_i+a})
\quad Mod(\frac{1}{(\th_i+t)^{n_i/2})}=((\th_i+a)^2+b^2)^{-n_i/4}
$$
$$
arg(e^{\frac{\gamma_i^2}{4(\th_i+t)}})=-\frac{\gamma_i^2}{4} \frac{b}{(\th_i+a)^2+b^2}\quad 
Mod(e^{\frac{\gamma_i^2}{4(\th_i+t)}})=e^{\frac{\gamma_i^2}{4} \frac{a+\th_i}{(\th_i+a)^2+b^2}}
$$
Therefore

$$\xi=Arg(e^{\frac{\gamma_i^2}{4(\th_i+t)}}{\prod_{i=1}^p(\th_i+t)^{\frac{-n_i}{2}} }   e ^{st})=-\sum_{i=1}^{p}\frac{n_i}{2}\tan^{-1}(\frac{b}{\th_i+a})+\frac{\gamma_i^2}{4} \frac{b}{(\th_i+a)^2+b^2}+sb$$

$$\rho=Mod(e^{\frac{\gamma_i^2}{4(\th_i+t)}}{\prod_{i=1}^p(\th_i+t)^{\frac{-n_i}{2}} }   e ^{st})=\sqrt{a^2+b^2}\prod_{i=1}^{p}((\th_i+a)^2+b^2)^{-n_i/4}e^{\frac{\gamma_i^2}{4} \frac{a+\th_i}{(\th_i+a)^2+b^2}}e^{sa} $$
The integration around the poles $-\th_i$ is the source of possible issues. Let us focus on a  typical situation where without loss of generality (due to the shift property) we can assume that the pole around which we will integrate along some contour $\Gamma_{\epsilon}=\{\epsilon e^{-i\xi}, \xi\in (0,2\pi)\}$}  

\fi 
\section{Concluding remarks}
In this paper we explore alternative expressions for the integral representation of this important family of distributions. It is clear that the branch cut approach covered here has simplicity in expression leading to an explicit connection between their pdf and df. Therefore one can apply the standard saddle point approximation method in evaluating tail probabilities.  
The resulting expressions in terms or branch cuts are also easy to implement using the standard numerical routines available in many computer packages. 
By identifying these cuts enables the user to vary the integrating contours and not just follow standard inversion methods or its improvements like Talbot methods. 
The saddle point approximation is shown to work well for a range of values, with a relative improvement in the tails of the distributions but such approximations are generally not performing well in the case of negative entries.


\bibliographystyle{elsarticle-harv} 

\bibliography{/Users/alfredkume/kent/research/biblio}
\appendix
\section{Proof of Theorem~1} 

In order to give now the idea of the proof, it is sufficient to consider as above the case $p=3$ while the generalization to larger $p$ easily follows. 
In order to see the derivation of the branch cuts first, let us assume for simplicity that $p=3$ and for a point $t$ in the complex plane just above the real line segment $[-\th_{2},-\th_{1}]$ we have that
$arg(\sqrt{(\th_1+t)}=r_1 e^{\text{\textbf{i}} \alpha_1 })=\alpha_{1}$ where $\alpha_{1}\approx \pi/2$ but $\alpha_{2}=arg(\sqrt{(\th_2+t)}\approx 0$ and similarly $\alpha_{3}\approx0$ and therefore $\alpha=\alpha_{1}+\alpha_{2}+\alpha_{3}\approx \pi/2$. However for a point just below $[-\th_{2},-\th_{1}]$ the same reasoning leads to 
$\alpha=\alpha_{1}+\alpha_{2}+\alpha_{3}\approx- \pi/2$. So  $\alpha\approx \pm\pi/2$ while taking different signs on either sides of $[-\th_{2},-\th_{1}]$ and hence the function $g(t)$ is not analytic in $[-\th_{2},-\th_{1}]$ implying  a branch cut. 
However, if we are to follow similar reasoning as above for points $t$ on either side of  $[-\th_{3},-\th_{2}]$ we then have $\alpha_{1}\approx\alpha_{2}\approx \pi/2$ and $\alpha_{3}\approx0$, i.e. $\alpha\approx \pi$ for the points just above and $\alpha\approx -\pi$ for the points just below this segment. Since the sign  choice for $\alpha\approx \pm\pi$ does not mater the function $g(t)$ remains analytic in $(-\th_{3},-\th_{2})$. While for the points around $[-\infty, -\th_{3}]$,  $\alpha_{1}\approx\alpha_{2} \approx\alpha_{3} \approx  \pm \pi/2$, i.e. $\alpha \approx \pm \frac{3}{2}\pi$ which suggests a branch cut. 

 \begin{figure}[ht]
	\begin{minipage}[b]{0.45\linewidth}
		\centering
		\centering
		\includegraphics[scale=.08]{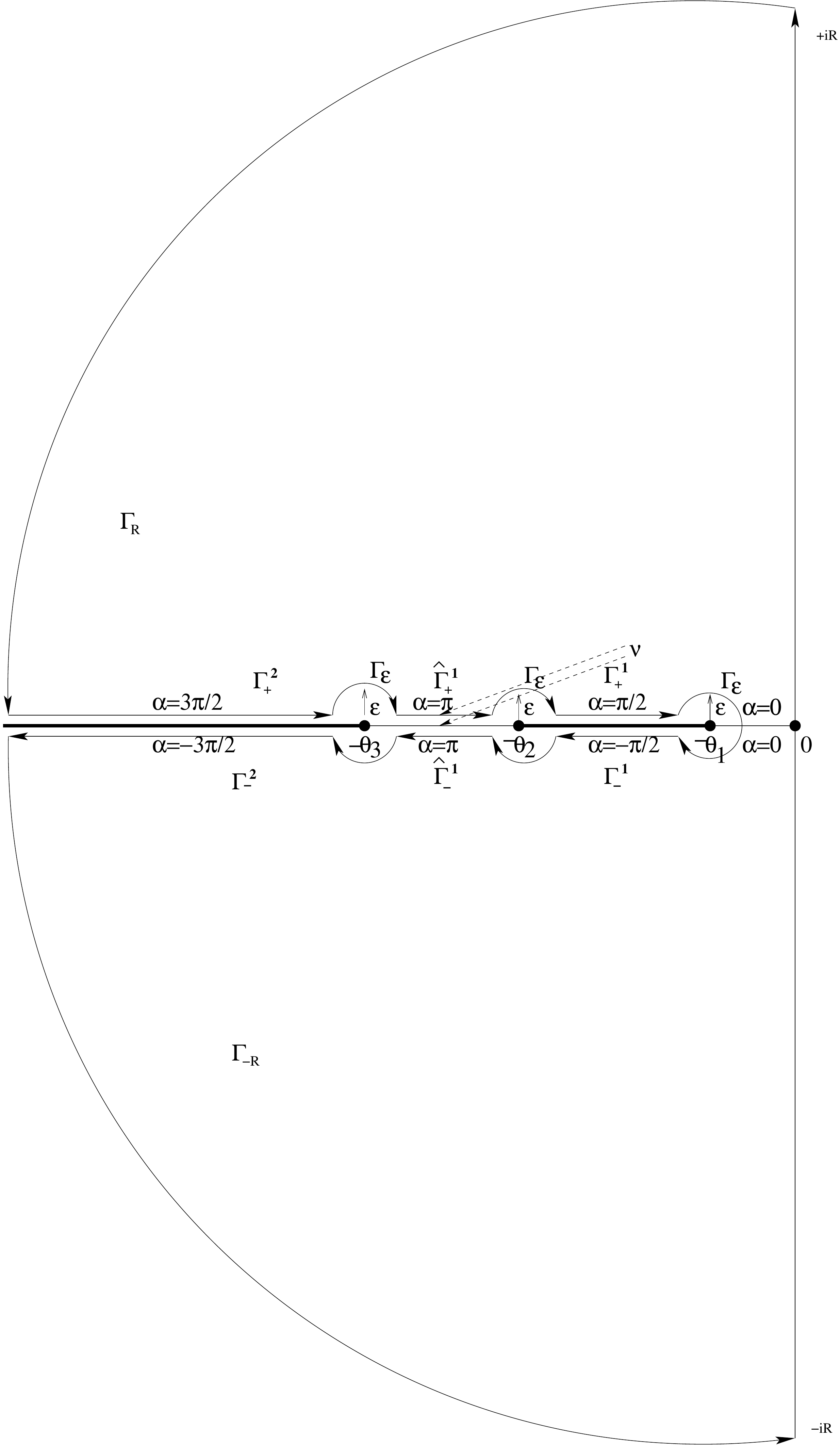} 
		\caption{Contour integration}
		\label{fig:cent}
	\end{minipage}
	\hspace{0.01cm}
	\begin{minipage}[b]{0.45\linewidth}
		\centering
		\includegraphics[scale=.08]{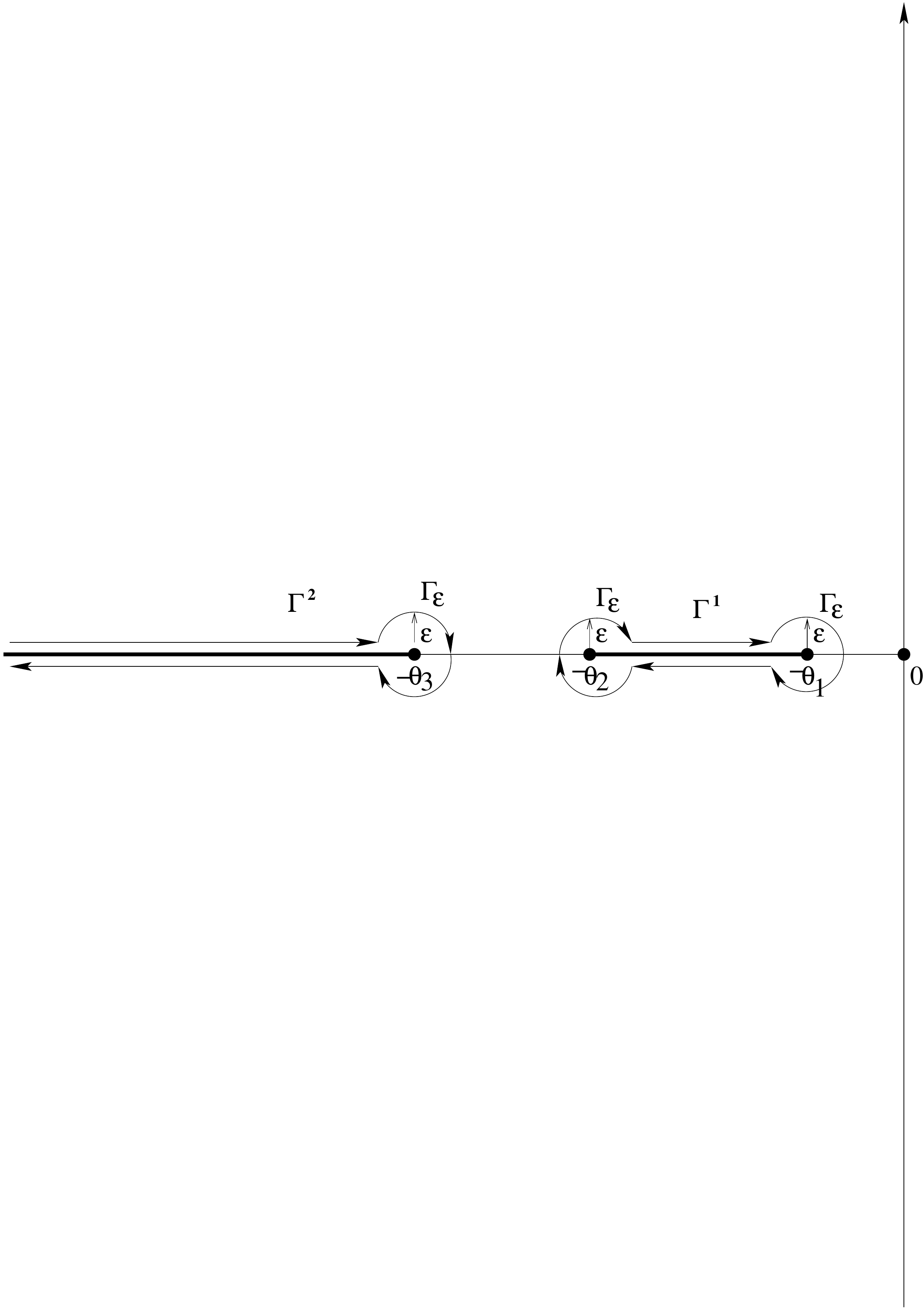} 
		\caption{Contour limits as $R\to \infty$ and $\nu\to0$ }
		\label{fig:mult}
	\end{minipage}
\end{figure}

We will focus now on the limiting behaviour of contours in Figure \eqref{fig:cent} as $R\to+\infty$ and  $\nu\to0$ while for the moment $\varepsilon$ is taken to be very small so that the regions containing the poles $\th_{i}$ resemble the ``dogbone'' or ``keyhole'' types as in Figure~\eqref{fig:mult}. 
The closed contour $\Gamma$ in Figure \eqref{fig:cent} is comprised of the vertical line $(-R, +R)$ and the elementary  parts $\Gamma_R, \Gamma_{-R}, \Gamma_\varepsilon, \Gamma_{\pm}, \hat{\Gamma}_{\pm}$. Applying the integration along the whole closed curve $\Gamma$, one can see that the function is analytic while it does not contain any poles and therefore the contour integration along this closed contour is zero. Now from theorem 4.3.2 in \cite{Ablowitz2003}  since the function $g(t) e^{st} $ is exponentially bounded i.e. $\lim_{t \to \infty}g(t) e^{st}  e^{\kappa}=0$ for some $\kappa$,  the curve integrals along the part-circle curves $\Gamma_R, \Gamma_{-R}$ disappear:
$$
\lim_{R \to \infty} \intl_{\Gamma_R  \cup \,\Gamma_{-R}} g(t) e^{st} dt =0$$
Since the integration on the regions $\hat{\Gamma}_+$ $\hat{\Gamma}_-$ is along the opposing directions it then follows that the integral components where the function values coincide cancel out:
$$
\lim_{\nu \to 0} \intl_{\hat{\Gamma}_+  \cup \, \hat{\Gamma}_{-}} g(t) e^{st} dt = 0
$$ 
Therefore the integration along the vertical line is  opposite the integration along the contours $\Gamma^{i}$. We can also easily see that without loss of generality if some of these contours circle some single point representing multiplicity of type 1.  

For deriving the second result of Theorem 1, one can notice that the survival function $P(X>x)$ is evaluated as 
\[
P(X>x)= \int_x^{\infty} f_{\mybth,\mybg^2}(s) ds=\kappa \left(\frac{1}{2 \pi\textbf{i}} \sum_{i=1}^{r} \intl_{\Gamma^{i}}   \frac{g(t)}{t} e^{tx}  dt +\frac{1}{2 \pi\textbf{i}} \sum_{j=1}^{l} \intl_{\Gamma^{j}_{\varepsilon}}   \frac{g(t)}{t} e^{tx}  dt\right)
\]
which is obtained by replacing the expression for $f_{\mybth,\mybg^2}(s)$ as in the first equation of Theorem 1 while $\frac{1}{t}$ appears by changing the order of integration such that $\int_{x}^{\infty} g(t) e^{ts}ds=-\frac{g(t)}{t} e^{xt}$ for complex $t$ such that $\text{Re}(t)<0.$ 

Note the fact that the integrand $\frac{g(t)}{t}$ has a simple pole at $0$ with residue $\kappa^{-1}=\frac{e^{\sum_{i=1}^p \frac{\gamma_i^2}{4\th_i}}}{ \prod_{i=1}^p\th_i^{\frac{n_i}{2}} }$. The original  integrating contour was the vertical line $\textbf{i}\R+t_0$ at some $-min(\mybth)<t_0<0$ which was then reduced to the elementary contours $\Gamma_i$ as in Theorem 1.

 Let us chose another integrating contour by changing $t_0$ to some  $t'_0>0$. Now  $\textbf{i}\R+t'_0$ will be on the right of the pole $0$  and the residue  of $\frac{g(t)}{t}$ at $t=0$ is the difference between these two integrals:
\begin{eqnarray}
\frac{1}{ 2\pi \textbf{i}}\intl_{\textbf{i}\R+t'_0}  \frac{g(t)}{t}   e ^{st} dt&-&\frac{1}{ 2\pi \textbf{i}}\intl_{\textbf{i}\R+t_0}  \frac{g(t)}{t}   e ^{st} dt=\kappa^{-1}
\nonumber
\\
\frac{\kappa}{ 2\pi \textbf{i}}\intl_{\textbf{i}\R+t'_0}  \frac{g(t)}{t}   e ^{st} dt&=&1+\frac{\kappa}{ 2\pi \textbf{i}}\intl_{\textbf{i}\R+t_0}  \frac{g(t)}{t}   e ^{st} dt.
\label{eqn:property:scale:shift:dist:new}
\end{eqnarray}
The appearance of the additional pole at $0$ can be seen as taking some particular $\th_i$ to be $0$ and with multiplicity $n_i=2$.  In fact, one can easily apply a shift parametrisation such that the pole $0$ is shifted to some $-\th_0$:
\[
\frac{1}{ 2\pi \textbf{i}}\intl_{\textbf{i}\R+t'_0}  \frac{g(t)}{t}   e ^{st} dt= e ^{s\th_0} \frac{1}{ 2\pi \textbf{i}}\intl_{\textbf{i}\R+t'_0-\th_0}  \frac{g(t+\th_0)}{t+\th_0}   e ^{st} dt=f_{\tilde{\mybth}, \kbm{0}}(x) e^{x\th_0} \frac{ \kappa}{\kappa'}\quad \text{with} \quad \kappa'=\frac{ \th_0 \prod_{i=1}^p\th_i^{\frac{n_i}{2}} }{e^{\sum_{i=1}^p \frac{\gamma_i^2}{4(\th_i-\th_0)}}}
\]
While noting that for $t_0<0$, $$P(X\le x)=1-P(X>x)=1-\frac{\kappa}{ 2\pi \textbf{i}} \int_x^{+\infty} \intl_{\textbf{i}\R+t_0}g(t)   e ^{st} dt ds= 1+\frac{\kappa}{ 2\pi \textbf{i}} \intl_{\textbf{i}\R+t_0}g(t) \frac{e ^{xt}}{t}dt$$ 
identity \eqref{eqn:property:scale:shift:dist:new} implies for the distribution function
\begin{eqnarray}
P(X\le x)=1- \int_x^{+\infty} f_{\mybth,\mybg^2}(s) ds
=\frac{e ^{x\th_0}}{x \th_0} f_{x\tilde{\mybth},x\tilde{\mybg}^2}(1) \prod_{i=1}^{p}\frac{\th_i^{\frac{n_i}{2}}}{(\theta_i+\th_0)^{\frac{n_i}{2}}}
e^{\sum_{i=1}^{p}\frac{\gamma^2_i}{4(\th_i+\th_0)}-\frac{\gamma^2_i}{4(\th_i)}}
\label{eqn:dist}
\end{eqnarray}
where $\tilde{\mybth}=(0,0,\mybth) +\th_0$ and $\tilde{\mybg}=(0,0,\mybg)$.

The second result follows from the discussion following equation \eqref{eqn:property:scale:shift:dist:new} whose derivation is  now easily confirmed since the integration along the unbounded contours $\textbf{i}\R+t_0$ and $\textbf{i}\R+t'_0$ is reduced to the elementary contours $\Gamma_i$ and $\Gamma_{\epsilon}^j$ as described earlier in the proof the theorem.

\endproof


\section{Proof of Theorem \ref{Theo:Bingh}}

Note now some important observations related to the contour integration around a simple pole.  Theorem 4.3.1 of \cite{Ablowitz2003} states that if at a particular pole $-\th$,  \AK{$g(t)(t+\th) \to 0$ as $t\to -\th_i$} then\footnote{If $\th$ is a pole of finite order $n+1$ then the residue theorem confirms  $$\lim_{\varepsilon\to 0}\frac{1}{2 \pi \textbf{i}}\int_{\Gamma_{\varepsilon}} g(t) dt=\frac{1}{n!} \frac{d^{n}}{d^{n}t}\left( g(\th)(t-\th)^{n} \right)$$
} 
$$\lim_{\varepsilon\to 0}\int_{\Gamma_{\varepsilon}} g(t) dt=0$$
In our case, since $\th_i$ have multiplicity 1 the term in $g(t)$ related to $\th_i$ is $\frac{1}{(\th_i+t)^{\frac{1}{2}}}$ and therefore $g(t)(t+\th)\to 0$  as $t\to -\th_i$ and so $\Gamma_\varepsilon$ contours around the poles are zero and in addition:
	$$\intl_{\Gamma_\varepsilon} g(t) e^{st} dt \to 0 \quad \nu,\varepsilon\to 0 $$
	where the integration is applied at each $\Gamma_\varepsilon$ around $-\th_{i}$ and $\nu$ here represents the distance of the horizontal parts of $\Gamma^{i}$ from the real axis (see Figure~\ref{fig:simple:gen})
	. This implies that  the contours $\Gamma^{i}$ as in Figure~\ref{fig:mult} could be  simplified to interval parts of real axis while poles having residue zero.
This limit holds since at each pole $-\theta_{i}$, we have multiplicity 1 and therefore $g(t) e^{st} (t+\th_i)\to 0$ as $t\to -\th_i$.
As a result the Bromwich contour integral along the vertical line is the opposite of the value of the contour integration along both horizontal contours which are on either sides of the real axis of  negative values. Now as $\nu\to 0$ both sides of $\Gamma^{i}$'s can be taken as close as possible to the real line until they both align with each other. 

However, while the branch cut regions $C_{i}$, (bolded line regions in the Figure~\ref{fig:simple:lim}) are the limits of $\Gamma^{i}$ as $\nu\to 0$, the argument values $\alpha$ of the function $\frac{g(t)}{2 \pi \mathbf{i}} e^{t}$ are such that $\alpha\to \pm (2r-1) \frac{\pi}{2}$. And, more importantly as  $\nu\to 0$, the argument $\alpha$ approaches with opposing signs while the regions $\Gamma_{\pm}$ go to $C_{r}$ . For example, $C_{1}$ in the figure  is the limit of $\Gamma^{1}_+$, as $\nu\to0$, but $\alpha$ approaches the value $\frac{\pi}{2}$ while in the region $\Gamma^{1}_-$, $\alpha$ approaches $-\frac{\pi}{2}$. This feature repeats itself in every branch cut region $C_{r}$ such that at regions $\Gamma_+$ $\Gamma_-$,   $\alpha$  becomes either $r \frac{\pi}{2}$ or $-r \frac{\pi}{2}$ respectively. While in such regions the limiting behavior of $g(t) e^{st}$ has a sign change this is then reversed by the opposing integration directions. This implies that the integration along the branch cut is doubled while in the other parts or the real axis is zero. Therefore after converting the elementary integration in real coordinates and by introducing $-1$ in the square root the result follows.

\endproof

	\section{Proof of Theorem 3}
The density function of $Z$ for $z\ge 0$  is
\begin{eqnarray*}
	f_Z(z)&=&\int_{0}^{\infty}f_{\mybth, \mybg^2}(y+z)f_{\mybth', \mybg'^2}(y) dy\\
	&=& \kappa \kappa'\int_{0}^{\infty}  \intl_{\textbf{i}\R+t_0} \intl_{\textbf{i}\R+v_0 }   e ^{(y+z)t} g(t)   e^{yv} g'(v) dv  dt dy\\
	&=&  \kappa \kappa' \intl_{\textbf{i}\R+t_0}  \intl_{\textbf{i}\R+v_0 }  e^{zt}  g(t) dt  g'(v) dv \int_{0}^{\infty} e ^{y(t+v)} dy  \\
	&=&  \frac{\kappa}{ 2\pi \textbf{i}}\frac{\kappa'}{{ 2\pi \textbf{i}}} \intl_{\textbf{i}\R+t_0}  \intl_{\textbf{i}\R+v_0 }  e^{zt}  g(t) dt  g'(v) dv \frac{-1}{t+v} \\
	&=&  \kappa \kappa' \frac{1}{ 2\pi \textbf{i}}\intl_{\textbf{i}\R+t_0}  e^{zt}  g(t)   g'(-t) dt
\end{eqnarray*}
in the identities above  we used the fact that $\int_{0}^{\infty} e ^{y(t+v)} dy=-\frac{1}{t+v}$ for complex numbers $t$ and $v$ which are both on the left of the line $\textbf{i}\R$ and from the residue theorem  $\frac{1}{ 2\pi \textbf{i}}\intl_{\textbf{i}\R+v_0 }  \frac{1}{t+v} g'(v) dv=-g'(-t)$ for any $t$ on the right of $\textbf{i}\R+v_0 $ (or the poles of $g(t)$). 
As in the previous cases of Theorems 1 and 2, the integrating contour could then be reduced to the poles or branch cuts on the left of the integrating line which involve only those poles related to $\kbm{\th}$. 
Additionally, by integrating on $(z,+\infty)$ one can obtain the right tail probability so that 
$$
P(Z>z>0)= \kappa \kappa' \frac{-1}{ 2\pi \textbf{i}}\intl_{\textbf{i}\R+t_0}  \frac{e^{zt}}{t}  g(t) g'(-t) dt
$$  
which has the same effect as in Theorem 1 case where the additional pole at $t=0$ on the left of the integrating contour leads to 
$$
\kappa \kappa' \frac{1}{ 2\pi \textbf{i}}\intl_{\textbf{i}\R+t_0}  \frac{e^{zt}}{t}  g(t) g'(-t) dt+\kappa \kappa' \frac{1}{ 2\pi \textbf{i}}\intl_{\textbf{i}\R+t'_0}  \frac{e^{zt}}{t}  g(t) g'(-t) dt=1
$$
for some $0<t'_0<min(\kbm{\th'})$.
As a result for $z>0$
$$
P(Z<z)=\frac{e^{\th_0 z}}{\th_0}f_{\tilde{Z}}(z)\prod_{i=1}^p\frac{\sqrt{\th_i}}{\sqrt{\th_i+\th_0}} \prod_{i=1}^{p'}\frac{\sqrt{\th'_i}}{\sqrt{\th'_i-\th_0}} e^{\sum_{i=1}^{p}\frac{\gamma^2_i}{4\th_i}-\frac{\gamma^2_i}{4(\th_i+\th_0)}} e^{\sum_{i=1}^{p'}\frac{\gamma'^2_i }{4\th'_i }-\frac{\gamma'^2_i }{4(\th'_i-\th_0) }}
$$
where  $$\tilde{Z}=\frac{1}{2\theta_0} \chi^2_2(0)+\sum_{i=1}^{p}
\frac{1}{2(\theta_i+\theta_0)} \chi^2_{n_i}(\frac{\gamma_i^2}{2 (\th_i+\th_0)})-\sum_{i=1}^{p'}
\frac{1}{2(\theta'_i-\theta_0)} \chi^2_{n'_i}(\frac{\gamma_i'^2}{2 (\th'_i-\th_0)})$$
with $0<\th_0<min(\kbm{\th}')$.
Noting that for $-Z=Y-X$, a  similar argument as above can be applied by switching the roles of $g$ and $g'$ so that for $z<0$ we obtain
$$
f_z(z)=\kappa \kappa' \intl_{\textbf{i}\R+v_0}  e^{-zt}  g(-t)   g'(t) dt \quad z<0
$$ 
\endproof

\end{document}